\documentclass[aps,pra,showpacs,amsmath,amssymb,amsfonts,superscriptaddress,twocolumn,lengthcheck]{revtex4-2}
\usepackage [latin1]{inputenc}
\usepackage{amsmath}
\usepackage{amsfonts}
\usepackage{graphicx}
\usepackage{color}
\usepackage[colorlinks,citecolor=blue]{hyperref}
\begin{document}

\title{Optimal control of linear Gaussian quantum systems via quantum learning control}
\author{Yu-Hong Liu}
\affiliation{Key Laboratory of Low-Dimensional Quantum Structures and Quantum Control of
Ministry of Education, Key Laboratory for Matter Microstructure and Function of Hunan Province, Department of Physics and Synergetic Innovation Center for Quantum Effects and Applications, Hunan Normal University, Changsha, 410081, China}
\author{Yexiong Zeng}
\affiliation{Theoretical Quantum Physics Laboratory, Cluster for Pioneering Research, RIKEN, Wakoshi, Saitama 351-0198, Japan}
\affiliation{Quantum Computing Center, RIKEN, Wakoshi, Saitama 351-0198, Japan}
\author{Qing-Shou Tan}
\affiliation{Key Laboratory of Hunan Province on Information Photonics and Freespace Optical Communication, College of Physics and Electronics, Hunan Institute of Science and Technology, Yueyang 414000, China}
\author{ Daoyi Dong}
\affiliation{School of Engineering and Information Technology, University of New South Wales, Canberra Australian Capital Territory 2600, Australia}
\author{Franco Nori}
\affiliation{Theoretical Quantum Physics Laboratory, Cluster for Pioneering Research, RIKEN, Wakoshi, Saitama 351-0198, Japan}
\affiliation{Quantum Computing Center, RIKEN, Wakoshi, Saitama 351-0198, Japan}
\affiliation{Department of Physics, The University of Michigan, Ann Arbor, Michigan, 48109-1040, USA}
\author{Jie-Qiao Liao}
\email{Corresponding author: jqliao@hunnu.edu.cn}
\affiliation{Key Laboratory of Low-Dimensional Quantum Structures and Quantum Control of
Ministry of Education, Key Laboratory for Matter Microstructure and Function of Hunan Province, Department of Physics and Synergetic Innovation Center for Quantum Effects and Applications, Hunan Normal University, Changsha, 410081, China}
\affiliation{Institute of Interdisciplinary Studies, Hunan Normal University, Changsha, 410081, China}

\begin{abstract}
Efficiently controlling linear Gaussian quantum (LGQ) systems is a significant task in both the study of fundamental quantum theory and the development of modern quantum technology. Here, we propose a general quantum-learning-control method for optimally controlling LGQ systems based on the gradient-descent algorithm. Our approach flexibly designs the loss function for diverse tasks by utilizing first- and second-order moments that completely describe the quantum state of LGQ systems. We demonstrate both deep optomechanical cooling and large optomechanical entanglement using this approach. Our approach enables the fast and deep ground-state cooling of a mechanical resonator within a short time, surpassing the limitations of sideband cooling in the continuous-wave driven strong-coupling regime. Furthermore, optomechanical entanglement could be generated remarkably fast and surpass several times the corresponding steady-state entanglement, even when the thermal phonon occupation reaches one hundred.~This work will not only broaden the application of quantum learning control, but also open an avenue for optimal control of LGQ systems.
\end{abstract}
\date{\today}
\maketitle

\section{INTRODUCTION}
Linear Gaussian quantum (LGQ) systems~\cite{WisemanPRL2005,Wiseman2009,Nurdin2017,Zhang2022}, characterized by both linear dynamics and Gaussian characteristic functions~\cite{WeedbrookRMP2012}, play a crucial role in quantum physics~\cite{Peres2002}, quantum control theory~\cite{Alessandro2007}, and quantum information science~\cite{Braunstein2005,WeedbrookRMP2012,Wangpr2007}. The LGQ systems can be used to describe various bosonic-mode systems, including optomechanical systems~\cite{Schwab2012PT,Kippenberg2014RMP,Bowen2016}, optical fields~\cite{WeedbrookRMP2012,Mandel1965}, atomic ensembles~\cite{Lukin2003,HammererRMP2010}, and Bose-Einstein condensates~\cite{Dalfovo1999,Morsch2006}. In fact, the LGQ systems represent a kind of important and typical quantum systems in physics. Consequently, the optimal control of LGQ systems is essential for both deepening the understanding of fundamental quantum physics~\cite{Peres2002} and promoting the development of quantum technology~\cite{BlochNJP2018,Benyoucef2023}.

Currently, several methods have been proposed to effectively control LGQ systems, such as quantum Kalman filtering~\cite{Kalman1960,Nurdin2017,Zhang2022}, quantum linear quadratic Gaussian control~\cite{Nurdin2017,Dong2010,Nurdin2009}, and feedback control~\cite{ZhangIEEETrans2011,Yamamotoprx2014,ZhangRP2017}. These methods have found applications in various optimization tasks, such as seeking for optimal unravellings~\cite{WisemanPRL2005}, optimizing quantum state elimination~\cite{WieczorekPRL2015},  enhancing entanglement~\cite{ManciniPRA2007,SerafiniPRL2010} and cooling effects~\cite{DohertyPRA1999,MagriniNature2021,Conangla2019,RosenzweigNature2021}, and deterministic quantum state preparation~\cite{Lukin2004}.
Recently, quantum learning control~\cite{Rabitz2000sc,Dong2020,JudsonPRL1992} has become a powerful approach for many complex
quantum control tasks, and it has been widely used in
molecular laser control and achieving superior control outcomes~\cite{JudsonPRL1992,TwamleyPRA2009,Zahedinejad2015,WuIE2017,NiuNPJ2019}. Nevertheless, the applicability of the quantum-learning-control method for optimal control of the LGQ systems remains unclear.

In this work, we develop a general quantum-learning-control method for LGQ systems to improve the output. Concretely, we apply the dynamic equations of first- and second-order moments to design the loss function for various control goals. These first- and second-order moments can completely reconstruct the quantum states of LGQ systems, which allows us to fully control the LGQ systems. We also give the gradient relationship between the control function and loss function using the variational chain rule. In particular, we apply the method to optimally control linearized optomechanical systems, which has received widespread attention due to
its scientific significance in the study of macroscopic quantum effects~\cite{Sekatski2018,LiaoPRL2016} and quantum precision measurements~\cite{CavesRMP1980,Giovannetti2004,MetcalfeAPR2014}. We derive how to achieve optimal optomechanical cooling of the mechanical resonator~\cite{LaiPRA2020,GiganN2006,ArcizetN2006,KlecknerN2006,Wilson-Rae2007PRL,Marquardt2007PRL,Dobrindt2008PRL,
Genes200877PRA,Chan2011Nature,Teufel2011Nature,Rossi2018Nature,Sommer2019PRL,LiuPRA2022} and optimal generation of optomechanical entanglement~\cite{Vitali2007PRL,Genes2008PRA,Palomaki2013Science,
Riedinger2016Nature,Ho2018PRL,KorppiN2018,Jiao2020PRL,Yu2020Nature,Lai2022PRL,RiedingerNA2018,Hong2017Sci}, which are two critical goals in cavity optomechanics.~Remarkably, our cooling results break the
single-cavity sideband-cooling limit, and the generated optomechanical entanglement exceeds several times the corresponding steady-state entanglement, demonstrating the great potential of our method in optimal control of LGQ systems.

The rest of this paper is organized as follows. In Sec.~\ref{sec2}, we introduce the LGQ systems and present the quantum-learning-control method. In Sec.~\ref{sec3}, we apply the quantum-learning-control method to the linearized optomechanical system and use it to optimize optomechanical cooling and optomechanical entanglement. In Sec.~\ref{sec4}, we present some discussions on the dynamics of both the displacement amplitude and normalized detuning, impact of laser amplitude and phase noises on the cooling and entanglement performances, deeper optomechanical cooling and larger optomechanical entanglement, and experimental implementation of our scheme. Finally, we conclude this work in Sec.~\ref{sec5}.

\section{Linear Gaussian quantum systems and quantum learning control\label{sec2}}
Consider an LGQ system consisting of $N$ bosonic modes, described by the quadrature operators
\begin{equation}
\mathbf{\hat{x}}=(\hat{q}%
_{1},\hat{p}_{1},\ldots,\hat{q}_{N},\hat{p}_{N})^{\mathrm{T}},
\end{equation}
where ``$\mathrm{T}$" denotes matrix transpose.~The quadrature operators obey the commutation relation ($\hbar =1$)
\begin{equation}
[\mathbf{\hat{x}}_{j},\mathbf{\hat{x}}_{k}]=i\mathbf{%
\Xi }_{jk},
\end{equation}
where $\mathbf{\Xi }$ is a $2N\times 2N$
symplectic matrix defined by the elements
\begin{equation}
\mathbf{\Xi }_{jk}=\frac{1}{2}\delta _{j+1,k}[1-(-1)^{j}]-\frac{1}{2}\delta
_{j,k+1}[1+(-1)^{j}],
\end{equation}
with the Kronecker delta function $\delta _{jk}$. The state of the LGQ system can be described by both the mean vector
\begin{equation}
\langle \mathbf{\hat{x}}\rangle =\mathrm{Tr}(\mathbf{\hat{x}}\hat{\rho} ),
\end{equation}
and the covariance matrix
\begin{equation}
\mathbf{V}=\frac{1}{2}\mathrm{Tr}[\{\mathbf{\hat{x}}-\langle\mathbf{\hat{x}}\rangle,(\mathbf{\hat{x}}-\langle\mathbf{\hat{x}}\rangle)^{\mathrm{T}}\}\hat{\rho}],
\end{equation}
related to the first- and second-order moments respectively, where $\hat{\rho} $ is the density operator of the system and  $\{\cdot,\cdot\}$ denotes the anticommutator.

We focus on the dynamics of the LGQ system governed by a general Hamiltonian with linear and quadratic terms~\cite{WisemanPRL2005}
\begin{equation}
\hat{H}(t)=\frac{1}{2}\mathbf{\hat{x}}^{\mathrm{T}}\mathbf{G}(t)\mathbf{\hat{x}}-%
\mathbf{\hat{x}}^{\mathrm{T}}\mathbf{\Xi }\mathbf{\ u}(t),  \label{eq1}
\end{equation}%
where $\mathbf{G}(t)$ is a time-dependent $2N\times2N$ matrix, while $\mathbf{u}(t)$ is a $2N$-dimensional vector. Both $\mathbf{G}(t)$ and $\mathbf{u}(t)$ are the control parameters.~Including weak dissipations, the evolution of the system can be governed by the Lindblad master equation
\begin{equation}
\dot{\hat{\rho}}=-i[\hat{H}(t),\hat{\rho} ]+\mathcal{D}%
( \hat{\rho} ),\label{master}
\end{equation}
where $\mathcal{D}( \hat{\rho} ) \equiv
\sum_{j=1}^{M}(\hat{L}_{j}\hat{\rho} \hat{L}_{j}^{\dagger }-\{\hat{L}_{j}^{\dagger
}\hat{L}_{j},\rho \}/2)$,
with  $\hat{L}_{j}=
\sum_{k=1}^{2N}( \mathbf{d}_{j}) _{k}\hat{\mathbf{x}}_{k}$ ($\mathbf{d}_{j}$ denotes $2N$-dimensional complex vectors). Based on the quantum master equation~(\ref{master}) and the relation
\begin{eqnarray}
\frac{d}{dt}\langle \hat{O}\rangle =\mathrm{Tr}[\hat{O}\dot{\hat{\rho}}],
\end{eqnarray}
we can obtain the derivatives of the first-order moment and covariance matrix as
\begin{subequations}
\begin{eqnarray}
\frac{d\langle \mathbf{\hat{x}}( t) \rangle}{dt}  &=&\mathbf{A}( t) \langle \mathbf{\hat{x}}( t) \rangle +\mathbf{u}( t) ,  \label{eq21} \\
\frac{d\mathbf{V}( t)}{dt} &=&\mathbf{A}( t) \mathbf{V}( t)+%
\mathbf{V}( t) \mathbf{A}^{\mathrm{T}}( t) +\mathbf{E},
\label{eq22}
\end{eqnarray}%
\end{subequations}
where we introduce the drift matrix
\begin{equation}
\mathbf{A}(t)=\mathbf{\Xi }[\mathbf{G}(t)+\mathrm{Im}[\mathbf{D}^{\dag
}\mathbf{D}]]
\end{equation}
and the diffusion matrix
\begin{equation}
\mathbf{E}=\mathbf{\Xi }\mathrm{Re}[\mathbf{D}^{\dag }%
\mathbf{D}]\mathbf{\Xi}^{\mathrm{T}},
\end{equation}
with $\mathbf{D}=(\mathbf{d}_{1},%
\mathbf{d}_{2},\ldots,\mathbf{d}_{M})^{\mathrm{T}}$ being an $M\times 2N$
complex dissipation matrix.

The primary goal in controlling the LGQ systems is to optimize their dynamics for specific tasks by iteratively adjusting the control parameters with the gradient-descent algorithm.~The core of this optimization is to minimize the loss function, denoted as $C(\langle \mathbf{\hat{x}}\rangle, \mathbf{V})$, which is a complex function of $\langle \mathbf{\hat{x}}\rangle$ and $\mathbf{V}$. Note that we can flexibly design the loss functions to meet the specific requirements of different control tasks. To this end, we use the gradient-descent algorithm
\begin{equation}
\mathcal{Q}_{l+1}=\mathcal{Q}_{l}-\chi _{\mathcal{Q}} \frac{\delta C(\langle \mathbf{\hat{x}}\rangle ,\mathbf{V})}{\delta \mathcal{Q}_{l}} %
,
\end{equation}
where $\mathcal{Q}$ represents the control parameters, $l$
indicates the iteration number, and $\chi _{\mathcal{Q}}$ is the learning
rate. The variation operation $\delta C(\langle \mathbf{\hat{x}}%
\rangle ,\mathbf{V}) /\delta \mathcal{Q}$ follows the chain rule
\begin{eqnarray}
\frac{\delta C(\langle \mathbf{\hat{x}}\rangle ,\mathbf{V})}{\delta
\mathcal{Q}} = \frac{\delta C( \langle \mathbf{\hat{x}}\rangle ,%
\mathbf{V})}{\delta \langle \mathbf{\hat{x}}\rangle}
\frac{\delta \langle \mathbf{\hat{x}}\rangle}{\delta \mathcal{Q}} +\frac{\delta C(\langle \mathbf{\hat{x}}\rangle ,\mathbf{V})}{\delta \mathbf{V}} \frac{\delta \mathbf{V}}{\delta \mathcal{Q}},
\end{eqnarray}%
where~$\delta C(\langle \mathbf{\hat{x}}\rangle ,\mathbf{V})
/\delta \langle \mathbf{\hat{x}}\rangle $ and $\delta C(\langle
\mathbf{\hat{x}}\rangle ,\mathbf{V})/\delta \mathbf{V}$ can be derived
from the explicit expression of $C(\langle
\mathbf{\hat{x}}\rangle ,\mathbf{V})$, while $\delta \langle \mathbf{\hat{x}}\rangle /\delta \mathcal{Q}$ and $\delta\mathbf{V} /\delta \mathcal{Q}$ are obtained by taking variation for both sides of Eqs.~(\ref{eq21}) and~(\ref{eq22}). A minimization of $C(\langle
\mathbf{\hat{x}}\rangle ,\mathbf{V})$ is achieved with the optimization method when a satisfactory result is obtained.

For the first-order moment $\langle \mathbf{%
\hat{x}}\rangle $, we take the variation with respect to $\mathcal{Q}(s)$ for both sides
of Eq.~(\ref{eq21}) and obtain
\begin{equation}
\frac{\delta \langle \mathbf{\dot{\hat{x}}}(t)\rangle }{\delta \mathcal{Q}(s)%
}=\mathbf{A}(t)\frac{\delta \langle \mathbf{\hat{x}}(t)\rangle }{\delta
\mathcal{Q}(s)}+\frac{\delta \mathbf{A}(t)}{\delta \mathcal{Q}(s)}\langle
\mathbf{\hat{x}}(t)\rangle +\frac{\delta \mathbf{u}(t)}{\delta \mathcal{Q}(s)%
}.  \label{s122eq}
\end{equation}%
Since $\mathcal{Q}(s)$ is a function of $s$, then $\frac{\delta \langle
\mathbf{\dot{\hat{x}}}(t)\rangle }{\delta \mathcal{Q}(s)}$ can be changed to
$\frac{d}{dt}\frac{\delta \langle \mathbf{\hat{x}}(t)\rangle }{\delta
\mathcal{Q}(s)}$~\cite{Greiner1996sp}. For the initial condition $\langle \mathbf{\hat{x}}(0)\rangle
=(C_{1},\ldots,C_{2N})_{2N\times 1}^{\text{T}}$, the variation of $\langle
\mathbf{\hat{x}}(0)\rangle $ with respect to $\mathcal{Q}(s)$ is
\begin{equation}
\frac{\delta \langle \mathbf{\hat{x}}(0)\rangle }{\delta \mathcal{Q}(s)}%
=(0,\ldots,0)_{2N\times 1}^{\text{T}},
\end{equation}%
and the solution of Eq.~(\ref{s122eq}) at the target time $T$
is given by
\begin{eqnarray}
\frac{\delta \langle \mathbf{\hat{x}}(T)\rangle }{\delta \mathcal{Q}(s)}
\!=\!\mathbf{U}(T)\mathbf{U}^{-1}(s)\left[ \frac{\delta \mathbf{A}(s)}{\delta
\mathcal{Q}(s)}\langle \mathbf{\hat{x}}(s)\rangle +\frac{\delta \mathbf{u}(s)%
}{\delta \mathcal{Q}(s)}\right] ,\label{supeqs7}
\end{eqnarray}
where $\mathbf{U}(t)$ satisfies $\dot{\mathbf{U}}(t)=\mathbf{A}(t)\mathbf{U}%
(t)$, with the initial identity matrix $\mathbf{U}(0)=\mathbf{I}_{2N\times2N}$. In the above calculations, we use the property $\delta\mathcal{Q}(\tau )/\delta \mathcal{Q}(s)=\delta(\tau-s)$~\cite{Greiner1996sp}. Note that the value
of $\delta \mathbf{A}(s)/\delta \mathcal{Q}(s)$ and $\delta \mathbf{u}%
(s)/\delta \mathcal{Q}(s)$ can be accurately obtained according to the
specific expressions of $\mathbf{A}(s)$ and $\mathbf{u}(s)$.

Further, we vectorize the covariance matrix $\mathbf{V}%
(t)$ to find the variation of $\mathbf{V}(t)$ versus $\mathcal{Q}(s)$~\cite%
{Henderson1981}. By vectorizing both sides of Eq.~(\ref{eq22}), we obtain
\begin{eqnarray}
\mathrm{vec}[\mathbf{\dot{V}}(t)] &=&\mathrm{vec}[\mathbf{A}(t)\mathbf{V}%
(t)]+\mathrm{vec}[\mathbf{V}(t)\mathbf{A}^{\mathrm{T}}(t)]+\mathrm{vec}(%
\mathbf{E)}  \notag \\
&=&[\mathbf{I}\otimes \mathbf{A}(t)+\mathbf{A}(t)\otimes \mathbf{I}]\mathrm{%
vec}[\mathbf{V}(t)]+\mathrm{vec}(\mathbf{E)}, \notag \\
\label{supeqs88}
\end{eqnarray}%
with
\begin{equation}
\mathrm{vec}[\mathbf{V}]=\left(
\mathbf{V}_{11},\ldots,\mathbf{V}_{1N},\ldots,\mathbf{V}_{2N2N}%
\right)^{\mathrm{T}} .
\end{equation}%
In the above calculations, we use the property $\mathrm{vec}(\mathbf{XYZ})=(%
\mathbf{Z}^{\text{T}}\bigotimes \mathbf{X})\mathrm{vec}(\mathbf{Y})$~\cite%
{Henderson1981}. To better understand Eq.~(\ref{supeqs88}), we define $%
\mathbf{J}(t)=\mathbf{I}\otimes \mathbf{A}(t)+\mathbf{A}(t)\otimes \mathbf{I}
$, $\mathbf{K}(t)=\mathrm{vec}[\mathbf{V}(t)]$, and $\mathbf{L}=\mathrm{vec}(%
\mathbf{E)}$, then we have
\begin{equation}
\mathbf{\dot{K}}(t)=\mathbf{J}(t)\mathbf{K}(t)+\mathbf{L},  \label{speq133}
\end{equation}%
which has the same form as Eq.~(\ref{eq21}). In the same way, we obtain%
\begin{equation}
\frac{d}{dt}\left[\frac{\delta \mathbf{K}(t)}{\delta \mathcal{Q}(s)}\right]=\mathbf{J}(t)\frac{%
\delta \mathbf{K}(t)}{\delta \mathcal{Q}(s)}+\frac{\delta \mathbf{J}(t)}{%
\delta \mathcal{Q}(s)}\mathbf{K}(t),  \label{supeq22}
\end{equation}%
and the solution of Eq.~(\ref{supeq22}) at the target time $T$ is%
\begin{equation}
\frac{\delta \mathbf{K}(T)}{\delta \mathcal{Q}(s)}=\mathbf{R}(T)\mathbf{R}%
^{-1}(s)\frac{\delta \mathbf{J}(s)}{\delta \mathcal{Q}(s)}\mathbf{K}(s),
\label{supeq321}
\end{equation}%
where $\mathbf{R}(t)$ satisfies $\mathbf{\dot{R}}(t)=\mathbf{J}(t)\mathbf{R}%
(t)$, with the initial identity matrix $\mathbf{R}(0)=\mathbf{I}^{\prime}_{4N^{2}\times 4N^{2}}$. The value of $\delta \mathbf{J}(s)/\delta \mathcal{Q}(s)$ can also be calculated with the
specific expressions of $\mathbf{J}(s)$. Equations~(\ref{supeqs7}) and~(\ref{supeq321}) give the variation relations for the elements of $\langle \mathbf{\hat{x}}(T)\rangle$ and $\mathbf{V}(T)$ with respect to the control function $\mathcal{Q}(s)$, and the loss function $C(\langle \mathbf{\hat{x}}\rangle, \mathbf{V})$ is a complex function of $\langle \mathbf{\hat{x}}(T)\rangle$ and $\mathbf{V}(T)$. Therefore, we establish the variational relation between the loss function and the control function, then the gradient-descent algorithm can be used to achieve iterative optimization.

\section{Application of the control method to optomechanical systems \label{sec3}}
As an application of the general quantum-learning-control theory, we consider optimal optomechanical cooling and entanglement in a cavity optomechanical system.
\subsection{Linearized optomechanical Hamiltonian and its quadrature representation}
In a rotating frame defined by the unitary operator $\exp(-i\omega_{\mathrm{L}}t\hat{a}^{\dagger}\hat{a})$ ($\omega_{%
\mathrm{L}}$ is the carrier frequency of the pulsed drive), the Hamiltonian
of the optomechanical system reads ($\hbar=1$)
\begin{eqnarray}
\hat{H}_{\mathrm{opt}}(t)&=&\Delta _{c}\hat{a}^{\dag }\hat{a}+\omega _{\mathrm{%
m}}\hat{b}^{\dag }\hat{b}-g_{0}\hat{a}^{\dag }\hat{a}( \hat{b}^{\dag }+\hat{b%
})\notag \\
&&+ \Omega(t)e^{-i\phi(t)}\hat{a}^{\dag }+\Omega(t)e^{i\phi(t)}\hat{a} ,
\label{s111}
\end{eqnarray}
where $\hat{a}$ ($\hat{a}^{\dagger}$) and $\hat{b}$ ($\hat{b}^{\dagger}$)
are the annihilation (creation) operators of the cavity field (with
resonance frequency $\omega_{\mathrm{c}}$) and the mechanical resonator ($%
\omega_{\mathrm{m}}$), respectively. Here, $\Delta _{\mathrm{c}}=\omega_{%
\mathrm{c}}-\omega_{\mathrm{L}}$ is the detuning of the cavity frequency $%
\omega_{\mathrm{c}}$ with respect to the carrier frequency $\omega_{\mathrm{L%
}}$ of the driving pulse. The $g_{0}$ term describes the optomechanical
coupling between the cavity field and the mechanical resonator, with $g_{0}$
being the single-photon optomechanical-coupling strength. The $%
\Omega(t)e^{\pm i\phi(t)}$ terms denote the pulsed drive of the cavity
field, with $\Omega(t)$ and $\phi(t)$ being the driving amplitude and phase, respectively.

For open quantum systems, we assume that the cavity field is connected to a vacuum bath, and the mechanical resonator is coupled to a heat bath. Considering the Markovian-dissipation case, the evolution of the system is governed by the following quantum master equation
\begin{eqnarray}
\dot{\hat{\rho}}(t)&=&i[\hat{\rho}(t),\hat{H}_{\mathrm{opt}}(t)]+\kappa%
\mathcal{D}[\hat{a}]\hat{\rho}(t)\notag \\ &&+\gamma_{\mathrm{m}}(\bar{n}_{\mathrm{m}}+1)%
\mathcal{D}[\hat{b}]\hat{\rho}(t)+ \gamma_{\mathrm{m}}\bar{n}_{\mathrm{m}}\mathcal{%
D}[\hat{b}^{\dagger}]\hat{\rho}(t) ,  \label{s1eq1}
\end{eqnarray}
where $\hat{\rho}(t)$ is the density matrix of the optomechanical system, $%
\hat{H}_{\mathrm{opt}}(t)$ is given by Eq.~(\ref{s111}), and $\mathcal{D}[%
\hat{o}]\hat{\rho}=\hat{o}\hat{\rho} \hat{o}^{\dag }-(\hat{o}^{\dag }\hat{o}%
\hat{\rho}+\hat{\rho} \hat{o}^{\dag }\hat{o})/2$ (for $\hat{o}=\hat{a}$, $%
\hat{a}^{\dag }$, $\hat{b}$, and $\hat{b}^{\dag }$) are the standard
Lindblad superoperators~\cite{Agarwal2013,Scully1997}. The parameters $%
\kappa $ and $\gamma _{\mathrm{m}}$ are the decay rates of the cavity field
and the mechanical resonator, respectively, and $\bar{n}_{\mathrm{m}}$ is
the environmental thermal-excitation occupation associated with the
mechanical resonator.

To eliminate the driving terms of the operators $\hat{a}$ and $\hat{b}$ (the
$g_{0}$ term can be understood as an effective drive to mode $\hat{b}$), we
make the displacement transformations $D_{a}(\alpha )=\exp (\alpha \hat{a}%
^{\dagger}-\alpha ^{\ast }\hat{a})$ and $D_{b}(\beta )=\exp (\beta \hat{b}%
^{\dagger }-\beta^{\ast }\hat{b})$ for the density operator $\hat{\rho}(t)$, namely,
\begin{equation}
\hat{\rho}^{\prime }(t)=D_{a}(\alpha )D_{b}(\beta )\hat{\rho}(t)
D_{b}^{\dagger }(\beta )D_{a}^{\dagger }(\alpha ) .  \label{s1eq2}
\end{equation}
Here, $\hat{\rho}^{\prime }(t)$ is the density matrix of the optomechanical
system in the displacement representation, $\alpha(t)$ and $\beta(t)$ are
the displacement amplitudes of the cavity field and the mechanical motion,
respectively. The values of $\alpha(t)$ and $\beta(t)$ are determined by the
equations of motion for the semiclassical motion. By substituting Eq.~(\ref{s1eq2}) into Eq.~(\ref{s1eq1}) and setting the coefficient of the driving
terms to be zero, we obtain the quantum master equation in the displacement
representation,
\begin{eqnarray}
\dot{\hat{\rho}}^{\prime }(t)&=&i[\hat{\rho} ^{\prime }(t),\hat{H}_{\mathrm{dis%
}}(t)]+\kappa \mathcal{D}[\hat{a}]\hat{\rho} ^{\prime }(t)
\notag \\&&\!+\!\gamma _{\mathrm{m%
}}(\bar{n}_{\mathrm{m}}+1)\mathcal{D}[\hat{b}]\hat{\rho} ^{\prime
}(t)\!+\!\gamma _{\mathrm{m}}\bar{n}_{\mathrm{m}}\mathcal{D}[\hat{b}^{\dagger }]%
\hat{\rho} ^{\prime }(t) ,  \label{s1eq3}
\end{eqnarray}%
where we introduce the Hamiltonian in the displacement representation as
\begin{eqnarray}
\hat{H}_{\mathrm{dis}}(t)&=&\Delta (t)\hat{a}^{\dagger }\hat{a}-g_{0}\hat{a}%
^{\dagger }\hat{a}(\hat{b}+\hat{b}^{\dagger })+\omega _{\mathrm{m}}\hat{b}%
^{\dagger }\hat{b}\notag\\ &&+[G(t)\hat{a}^{\dagger }+G^{\ast }(t)\hat{a}](\hat{b}+\hat{%
b}^{\dagger }) .  \label{s1eq4}
\end{eqnarray}%
In Eq.~(\ref{s1eq4}), we introduce the normalized detuning $\Delta (t)
\equiv \Delta_{\mathrm{c}} +g_{0}[\beta(t)+\beta^{*}(t)]$ and the linearized
optomechanical coupling strength $G(t)\equiv g_{0}\alpha(t)$. The $\Delta (t)$ and $G(t)$ provide the physical feature of the pulsed optomechanics~\cite{Hong2017Sci,RiedingerNA2018,MacCabe2020Sc,MariPRL2009,VannerPNAS2011,HoferPRA2011,LiaoPRA201184,MachnesPRL2012,VannerNC2013,MeenehanPRX2015,Palomaki2013,FedoseevPRL2021}.

We consider
the strong-driving case such that $|\alpha(t)|\gg1$, then the trilinear
term (i.e., the $g_{0}$ term) in Eq.~(\ref{s1eq4}) can be safely omitted. As a result, we obtain
the linearized optomechanical Hamiltonian
\begin{equation}
\hat{H}_{\mathrm{lin}}(t)=\Delta (t)\hat{a}^{\dagger }\hat{a}+\omega _{%
\mathrm{m}}\hat{b}^{\dagger }\hat{b}+[G(t)\hat{a}^{\dagger }+G^{\ast }(t)%
\hat{a}](\hat{b}^{\dagger }+\hat{b}).  \label{seq1}
\end{equation}
Mathematically, the linearization Hamiltonian $\hat{H}_{\mathrm{lin}}(t)$ is bilinear, and hence
both the equations of motion for the first- and second-order moments will be closed.

To better apply the quantum-learning-control method, we transform the
Hamiltonian~(\ref{seq1}) into the quadrature representation, namely,
\begin{equation}
\hat{H}_{\mathrm{OM}}(t)\!=\!\frac{1}{2}\hat{\mathbf{x}}%
^{\mathrm{T}}_{\mathrm{OM}}\left(
\begin{array}{cc}
\Delta ( t) \mathbf{I} & \mathbf{S}(t) \\
\mathbf{S}^{\dag }(t) & \omega _{\mathrm{m}}\mathbf{I}%
\end{array}%
\right) \hat{\mathbf{x}}_{\mathrm{OM}}\!-\!\frac{\Delta (t) +\omega _{\mathrm{m}}}{2},
\label{eq1}
\end{equation}
where $\hat{\mathbf{x}}_{\mathrm{OM}}=(\hat{q}_{a},\hat{p}_{a},\hat{q}_{b},\hat{p}%
_{b})^{\mathrm{T}}$ is the quadrature operator vector with $\hat{q}_{o}=(\hat{o}%
^{\dag}+\hat{o})/\sqrt{2}$ and $\hat{p}_{o}=i(\hat{o}^{\dag }-\hat{o})/\sqrt{%
2}$ for $o=a,b$. The matrices $\mathbf{I}$ and $\mathbf{S}(t)$ are defined
as a $2\times2$ identity matrix and a $2\times2$ matrix with the form $%
[2G(t), 0 ; -2iG( t),0]$, respectively. Note that the term $[\Delta (t)
+\omega _{m}]/2$ can be ignored since it has no influence on the dynamics of the system,
merely introducing an energy shift.

To obtain the quantum master equation in the displacement representation, we
assume the coefficients of the driving terms to be zero. Then we obtain the
equations of motion for the displacement amplitudes $\alpha (t)$ and $\beta(t)$,
\begin{subequations}
\begin{eqnarray}
\dot{\alpha}(t)&=&-\left[i\Delta(t)+\frac{\kappa}{2}\right]\alpha(t)+i\Omega(t)\text{e}%
^{-i\phi(t)} ,  \label{eq6a} \\
\dot{\beta}(t)&=&-\left(i\omega_{\mathrm{m}}+\frac{\gamma_{\mathrm{m}}}{2}\right)\beta(t)-ig_{0}|\alpha(t)|^{2} .  \label{eq6b}
\end{eqnarray}
\end{subequations}
Particularly, Eqs.~(\ref{eq6a}) and~(\ref{eq6b}) determine the
relationship between the pulsed driving field [amplitude $\Omega(t)$ and
phase $\phi(t)$] and the displacement amplitudes [$\alpha(t)$ and $\beta(t)$%
], which indicates that the dynamic evolution of the system can be
controlled by adjusting the pulsed driving amplitude and phase.

Based on the Hamiltonian $\hat{H}_{\mathrm{OM}}(t)$, we can obtain the matrices $\mathbf{\ u}(t)$, $\mathbf{A}(t)$, and $\mathbf{E}$ for optomechanical systems. By solving Eqs.~(\ref{eq21}) and~(\ref{eq22}) with specified initial conditions, we obtain the dynamical evolution of $\langle \mathbf{\hat{x}}\rangle $ and $\mathbf{V}$. In modern experimental platforms, the linearized optomechanical system is mainly controlled by the driving field via $\mathbf{\ u}(t)$ and $\mathbf{A}(t)$. Therefore, we choose the optical drives as the control parameters, and the dynamic connection between the optical drives and $G(t)$ is determined by Eqs.~(\ref{eq6a}) and~(\ref{eq6b}). Below, we show the optimal control of optomechanical cooling and entanglement.

\subsection{Fast and deep optomechanical cooling}
In optomechanical systems, the cooling performance is characterized by the mean phonon number $\langle \hat{b}^{\dagger}\hat{b}\rangle$. For a given target time $T$, the smaller the value of $\langle \hat{b}^{\dagger}\hat{b}(T)\rangle$, the better the cooling performance. Therefore, $\langle \hat{b}^{\dagger}\hat{b}(T)\rangle$ is the loss function for the cooling problem. The mathematical expression of the gradient-descent algorithm for the
cooling process can be written as
\begin{equation}
\mathcal{Q}_{l+1}(s)=\mathcal{Q}_{l}(s)-\chi _{\mathcal{Q}} \frac{\delta \langle
\hat{b}^{\dag }\hat{b}(T)\rangle}{\delta \mathcal{Q}_{l}(s)}.
\end{equation}%
Considering the initial
condition $\langle \hat{a}(0)\rangle =$ $\langle \hat{a}^{\dagger
}(0)\rangle =$ $\langle \hat{b}(0)\rangle =$ $\langle \hat{b}^{\dagger
}(0)\rangle =0$, then all the elements of the first-order moment $\langle
\mathbf{\hat{x}}(s)\rangle $ are always zero. Therefore, we can re-express $%
\langle \hat{b}^{\dag }\hat{b}\rangle $ as a function of the covariance
matrix elements, defined as
\begin{equation}
\langle \hat{b}^{\dagger }\hat{b}\rangle =\frac{1
}{2}(\mathbf{V}_{33}+\mathbf{V}_{44}-1).
\end{equation}
The value of $\delta \langle
\hat{b}^{\dag }\hat{b}(T)\rangle /\delta \mathcal{Q}(s)$ can
be further written as%
\begin{equation}
\frac{\delta \langle \hat{b}^{\dag }\hat{b}(T)\rangle }{\delta
\mathcal{Q}(s)}=\frac{\delta \mathbf{V}_{33}(T)}{2\delta
\mathcal{Q}(s)}+\frac{\delta \mathbf{V}_{44}(T)}{2\delta
\mathcal{Q}(s)},
\end{equation}%
which is a function of $\delta \mathbf{V}_{33}(T)/\delta \mathcal{Q}(s)$ and $\delta \mathbf{V}_{44}(T)/\delta
\mathcal{Q}(s)$. The values of $\delta \mathbf{V}_{33}(T)/\delta \mathcal{Q}(s)$ and $\delta \mathbf{V}_{44}(T)/\delta
\mathcal{Q}(s)$ can be obtained based on Eq.~(\ref{supeq321}), but we need to find the matrix $\mathbf{J}(s)$ for the cavity optomechanical systems.

The complex dissipation matrix $\mathbf{D}$ for the optomechanical system can be written as
\begin{equation}
\mathbf{D}=\left(
\begin{array}{cccc}
\sqrt{\frac{\kappa }{2}} & i\sqrt{\frac{\kappa }{2}} & 0 & 0 \\
0 & 0 & \sqrt{\frac{\gamma _{\mathrm{m}}\left( \bar{n}_{\mathrm{m}}+1\right) }{2}} & i\sqrt{%
\frac{\gamma _{\mathrm{m}}\left( \bar{n}_{\mathrm{m}}+1\right) }{2}} \\
0 & 0 & \sqrt{\frac{\gamma _{\mathrm{m}}\bar{n}_{\mathrm{m}}}{2}} & -i\sqrt{\frac{\gamma _{\mathrm{m}}%
\bar{n}_{m}}{2}}%
\end{array}%
\right).
\end{equation}%
Then we can obtain
\begin{eqnarray}
\mathbf{E}&=&\frac{1}{2}\left(
\begin{array}{cccc}
\kappa  & 0 & 0 & 0 \\
0 & \kappa  & 0 & 0 \\
0 & 0 & \gamma _{\mathrm{m}}\left( 2\bar{n}_{\mathrm{m}}+1\right)  & 0 \\
0 & 0 & 0 & \gamma _{\mathrm{m}}\left( 2\bar{n}_{\mathrm{m}}+1\right)
\end{array}%
\right) .
\end{eqnarray}%
Combined with Hamiltonian (\ref{eq1}), we further obtain the
coefficient matrix
\begin{eqnarray}\label{eq32}
&\mathbf{A}(t)\!=\!\left(
\begin{array}{cccc}
-\frac{\kappa}{2} & \Delta (t) & -2iG(t) & 0 \\
-\Delta (t) & -\frac{\kappa}{2} & -2G(t) & 0 \\
0 & 0 & -\frac{\gamma _{\mathrm{m}}}{2} & \omega _{\mathrm{m}} \\
-2G^{\ast }(t) & -2iG^{\ast }( t)  & -\omega _{\mathrm{m}} & -\frac{\gamma _{\mathrm{m}}}{2}%
\end{array}%
\right).
\end{eqnarray}%
Using the relation $\mathbf{J}(t)=\mathbf{I}\otimes \mathbf{A}(t)+\mathbf{A}(t)\otimes \mathbf{I}$, we obtain the coefficient matrix $\mathbf{J}(t)$ for the optomechanical system as
\begin{equation}
\mathbf{J}(t)=\left(
\begin{array}{cc}
\mathbf{B} & \mathbf{C}%
\end{array}%
\right),\label{speq28}
\end{equation}%
with
\begin{widetext}
\small
\begin{subequations}
\begin{eqnarray}
\mathbf{B}&=&\left(
\begin{array}{cccccccc}
-\kappa  & \Delta (t) & -2iG(t) & 0 & \Delta (t) & 0 & 0 & 0 \\
-\Delta (t) & -\kappa  & -2G(t) & 0 & 0 & \Delta (t) & 0 & 0 \\
0 & 0 & -\left( \kappa +\gamma _{\mathrm{m}}\right)/2 & \omega _{\mathrm{m}} & 0 & 0 & \Delta
(t) & 0 \\
-2G^{\ast }(t) & -2iG^{\ast }\left( t\right)  & -\omega _{\mathrm{m}} & -\left(
\kappa +\gamma _{\mathrm{m}}\right)/2 & 0 & 0 & 0 & \Delta (t) \\
-\Delta (t) & 0 & 0 & 0 & -\kappa  & \Delta (t) & -2iG(t) & 0 \\
0 & -\Delta (t) & 0 & 0 & -\Delta (t) & -\kappa  & -2G(t) & 0 \\
0 & 0 & -\Delta (t) & 0 & 0 & 0 & -\left( \kappa +\gamma _{\mathrm{m}}\right)/2 &
\omega _{\mathrm{m}} \\
0 & 0 & 0 & -\Delta (t) & -2G^{\ast }(t) & -2iG^{\ast }\left( t\right)  &
-\omega _{\mathrm{m}} & -\left( \kappa +\gamma _{\mathrm{m}}\right)/2 \\
0 & 0 & 0 & 0 & 0 & 0 & 0 & 0 \\
0 & 0 & 0 & 0 & 0 & 0 & 0 & 0 \\
0 & 0 & 0 & 0 & 0 & 0 & 0 & 0 \\
0 & 0 & 0 & 0 & 0 & 0 & 0 & 0 \\
-2G^{\ast }(t) & 0 & 0 & 0 & -2iG^{\ast }\left( t\right)  & 0 & 0 & 0 \\
0 & -2G^{\ast }(t) & 0 & 0 & 0 & -2iG^{\ast }\left( t\right)  & 0 & 0 \\
0 & 0 & -2G^{\ast }(t) & 0 & 0 & 0 & -2iG^{\ast }\left( t\right)  & 0 \\
0 & 0 & 0 & -2G^{\ast }(t) & 0 & 0 & 0 & -2iG^{\ast }\left( t\right)
\end{array}%
\right),
\\
\mathbf{C}&=&\left(
\begin{array}{cccccccc}
-2iG(t) & 0 & 0 & 0 & 0 & 0 & 0 & 0 \\
0 & -2iG(t) & 0 & 0 & 0 & 0 & 0 & 0 \\
0 & 0 & -2iG(t) & 0 & 0 & 0 & 0 & 0 \\
0 & 0 & 0 & -2iG(t) & 0 & 0 & 0 & 0 \\
-2G(t) & 0 & 0 & 0 & 0 & 0 & 0 & 0 \\
0 & -2G(t) & 0 & 0 & 0 & 0 & 0 & 0 \\
0 & 0 & -2G(t) & 0 & 0 & 0 & 0 & 0 \\
0 & 0 & 0 & -2G(t) & 0 & 0 & 0 & 0 \\
-\left( \kappa +\gamma _{\mathrm{m}}\right)/2 & \Delta (t) & -2iG(t) & 0 & \omega
_{\mathrm{m}} & 0 & 0 & 0 \\
-\Delta (t) & -\left( \kappa +\gamma _{\mathrm{m}}\right)/2 & -2G(t) & 0 & 0 &
\omega _{\mathrm{m}} & 0 & 0 \\
0 & 0 & -\gamma _{\mathrm{m}} & \omega _{\mathrm{m}} & 0 & 0 & \omega _{\mathrm{m}} & 0 \\
-2G^{\ast }(t) & -2iG^{\ast }\left( t\right)  & -\omega _{\mathrm{m}} & -\gamma _{\mathrm{m}}
& 0 & 0 & 0 & \omega _{\mathrm{m}} \\
-\omega _{\mathrm{m}} & 0 & 0 & 0 & -\left( \kappa +\gamma _{\mathrm{m}}\right)/2 & \Delta (t)
& -2iG(t) & 0 \\
0 & -\omega _{\mathrm{m}} & 0 & 0 & -\Delta (t) & -\left( \kappa +\gamma _{\mathrm{m}}\right)/2 & -2G(t) & 0 \\
0 & 0 & -\omega _{\mathrm{m}} & 0 & 0 & 0 & -\gamma _{\mathrm{m}} & \omega _{\mathrm{m}} \\
0 & 0 & 0 & -\omega _{\mathrm{m}} & -2G^{\ast }(t) & -2iG^{\ast }\left( t\right)  &
-\omega _{\mathrm{m}} & -\gamma _{\mathrm{m}}%
\end{array}%
\right)\label{speq30}.
\end{eqnarray}
\end{subequations}
\end{widetext}
The coefficient matrix $\mathbf{J}(t)$ depends on the displacement amplitudes $\alpha(t)$ and $\beta(t)$  [$\Delta (t)
= \Delta_{\mathrm{c}} +g_{0}[\beta(t)+\beta^{*}(t)]$ and $G(t) = g_{0}\alpha(t)$], such that the variation of $\mathbf{J}(s)$ with respect to $\mathcal{Q}(s)$ is actually calculated by solving the variation of $\alpha(s)$ and $\beta(s)$ with respect to $\mathcal{Q}(s)$.

Taking the variation with respect to $\mathcal{Q}(s)$ on both sides of Eqs.~(\ref{eq6a})
and (\ref{eq6b}) as well as their Hermitian conjugate equations, we have~\cite{Greiner1996sp}
\begin{equation}
\dot{\mathbf{P}}(t)=\mathbf{W}(t)\mathbf{P}(t)+\mathbf{Q}(t),  \label{B2}
\end{equation}%
where
\begin{equation}
\mathbf{P}(t) =\left( \frac{\delta \alpha (t)}{\delta \mathcal{Q}(s)},\frac{%
\delta \beta (t)}{\delta \mathcal{Q}(s)},\frac{\delta \alpha ^{\ast }(t)}{%
\delta \mathcal{Q}(s)},\frac{\delta \beta ^{\ast }(t)}{\delta \mathcal{Q}(s)}%
\right) ^{\text{T}},
\end{equation}%
\begin{equation}
\small
\mathbf{W}(t)=\left(
\begin{array}{cccc}
-i\Delta (t)-\frac{\kappa }{2} & 0 & -ig_{0}\alpha(t) & -ig_{0}\alpha(t) \\
-ig_{0}\alpha ^{\ast }(t) & -ig_{0}\alpha(t) & -i\omega _{\mathrm{m}}-\frac{\gamma
_{\mathrm{m}}}{2} & 0 \\
0 & i\Delta (t)-\frac{\kappa }{2} & ig_{0}\alpha ^{\ast }(t) & ig_{0}\alpha
^{\ast }(t) \\
ig_{0}\alpha ^{\ast }(t) & ig_{0}\alpha(t) & 0 & i\omega _{\mathrm{m}}-\frac{%
\gamma _{\mathrm{m}}}{2}%
\end{array}%
\right) ,
\end{equation}%
and
\begin{equation}
\mathbf{Q}(t)=\left(
\begin{array}{c}
i\frac{\delta \Omega (t)}{\delta \mathcal{Q}(s)}e^{-i\phi (t)}+i\Omega (t)%
\frac{\delta \lbrack e^{-i\phi (t)}]}{\delta \mathcal{Q}(s)} \\
0 \\
-i\frac{\delta \Omega (t)}{\delta \mathcal{Q}(s)}e^{i\phi (t)}-i\Omega (t)%
\frac{\delta \lbrack e^{i\phi (t)}]}{\delta \mathcal{Q}(s)} \\
0%
\end{array}%
\right) .
\end{equation}%
The solution to Eq.~(\ref{B2}) is
\begin{equation}
\mathbf{P}(t)=\Lambda (t)\mathbf{P}(0)+\Lambda (t)\int_{0}^{t}\Lambda
^{-1}(\tau )\mathbf{Q}(\tau )d\tau ,  \label{B6}
\end{equation}%
where $\mathbf{P}(0)=(0,0,0,0)^{\text{T}}$, and $\Lambda (t)$ satisfies the
equation $\dot{\Lambda}(t)=\mathbf{W}(t)\Lambda (t)$ with the initial value
$\Lambda (0)=\mathbf{I}$. Then, we have
\begin{subequations}
\begin{align}
&\mathbf{P}_{\mathcal{Q}=\Omega}(s)=\left( \frac{%
1}{2}ie^{-i\phi (s)},0,-\frac{1}{2}ie^{i\phi (s)},0\right) ^{\text{T}}, \\
&\mathbf{P}_{\mathcal{Q}=\phi}(s)=\left( \frac{1%
}{2}\Omega (s)e^{-i\phi (s)},0,\frac{1}{2}\Omega (s)e^{i\phi (s)},0\right) ^{%
\text{T}}.
\end{align}%
\end{subequations}
Thus, we obtain the variation of the displacement amplitudes $\alpha (s)$
and $\beta (s)$ with respect to the driving amplitude $\Omega (s)$ and phase $\phi (s)$. The term $\delta \mathbf{K}(s)/\delta \mathcal{Q}(s)$ in Eq.~(\ref{supeq321}) gives the variation of all covariance-matrix elements with respect to $%
\mathcal{Q}(s)$, and the value of $\delta \langle b^{\dag }b(T)\rangle /\delta \mathcal{Q}(s)$ can be obtained. Based on the variation of $\langle \hat{b}^{\dagger}\hat{b}(T)\rangle$ with respect to the drive parameters $\Omega(s)$ and $\phi(s)$, the gradient-descent algorithm is executed until $\langle \hat{b}^{\dagger}\hat{b}(T)\rangle$ converges to a stable value.

\begin{figure}[tbp]
\centering \includegraphics[width=0.49\textwidth]{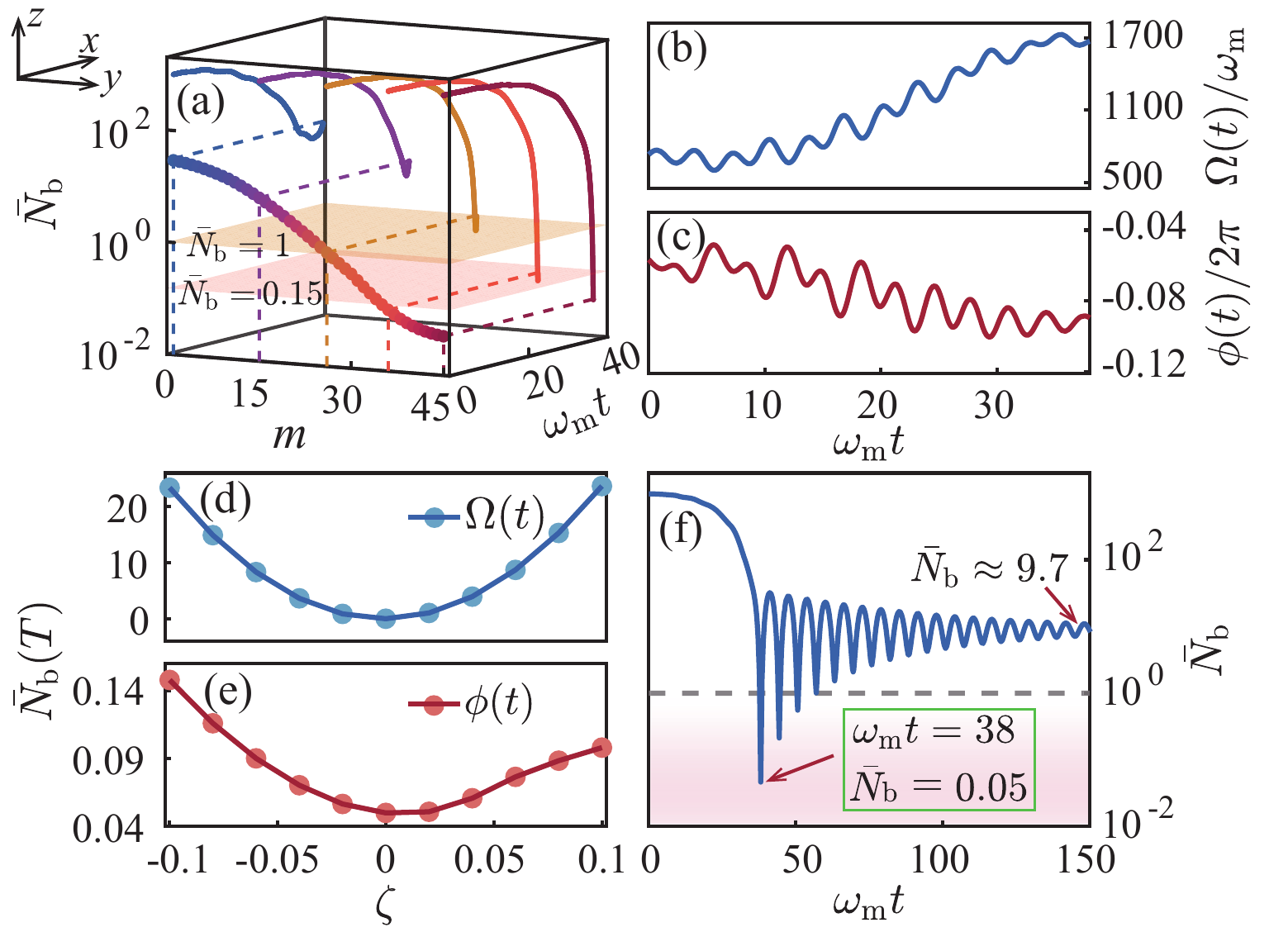}
\caption{(a) Mean phonon number $\bar{N}_{\text{b}}$ versus the scaled evolution time $\omega_{\mathrm{m}}t$ for different iteration numbers $m=0$, $15$, $26$, $36$, and $45$. The curve in the $yz$ plane shows $\bar{N}_{\text{b}}$ at the target time $T$ versus $m$. (b) Scaled driving amplitude $\Omega(t)/\omega_{\mathrm{m}}$ and (c) phase $\phi(t)/2\pi$ versus $\omega_{\mathrm{m}}t$, obtained after the last iteration ($m=45$). (d) The influence of the driving amplitude and (e) phase deviations on the cooling performance at $T$.~(f) Dynamic evolution of $\bar{N}_{\text{b}}$ after removing the pulsed drive. The gray dashed line marks the threshold for $\bar{N}_{\text{b}}=1$. Other parameters are $g_{0}/\omega_{\mathrm{m}}=4\times10^{-5}$, $\kappa/\omega_{\mathrm{m}}=0.02$, $\gamma_{\mathrm{m}}/\omega_{\mathrm{m}}=3\times10^{-6}$, $T=38\omega_{\mathrm{m}}^{-1}$, $\Delta_{\mathrm{c}}/\omega_{\mathrm{m}}=1$, and $\bar{n}_{\mathrm{m}}=10^{3}$. \label{fig1}}
\end{figure}
In Fig.~\ref{fig1}(a) we show the cooling process for different iteration numbers. Here we see that the increase of the iteration number will improve the cooling performance; for example, we have $\bar{N}_{\text{b}}=29.33$ for $m=0$ and $\bar{N}_{\text{b}}<1$ for $m=26$. In particular, the system enters the strong-coupling regime [$|G|>\{\kappa,\gamma_{\mathrm{m}}\} \rightarrow|\alpha|>\mathrm{max}\{\{\kappa,\gamma_{\mathrm{m}}\}/g_{0}\}=500$] in most of the duration. The cooling result ($\bar{N}_{\text{b}}=0.15$ for $m=36$) can break the single-cavity sideband-cooling limit ($n_{\mathrm{f}}=\bar{n}_{\mathrm{m}}\gamma_{\mathrm{m}}/\kappa\approx0.15$~\cite{Dobrindt2008PRL,He2017PRL} with $\bar{n}_{\mathrm{m}}$ being
the environmental thermal-excitation occupation associated with the
mechanical resonator) in the strong-coupling regime, and a deeper ground-state cooling of the mechanical mode can be realized ($\bar{N}_{\text{b}}=0.05$ for $m=45$). Note that we assume the mechanical resonator is initially in a thermal state with the same temperature as its heat bath. The density matrix of the thermal state is $\hat{\rho}_{\mathrm{th}}=\sum_{n=0}^{\infty}P_{n}|n\rangle\langle n|$, where $P_{n}=\bar{n}_{\mathrm{th}}^{n}/(\bar{n}_{\mathrm{th}}+1)^{n+1}$ is the probability distribution. The mean phonon number is given by $\bar{n}_{\mathrm{th}}=\langle \hat{n}\rangle=\mathrm{Tr}(\hat{\rho}_{\mathrm{th}}\hat{n})=1/(\mathrm{e}^{\hbar\omega_{\mathrm{m}}/k_{B}T}-1)$. This implies that the initial thermal state of the mechanical resonator can be characterized using the mean phonon number.

We display in Figs.~\ref{fig1}(b) and~\ref{fig1}(c) both the amplitude $\Omega(t)$ and phase $\phi(t)$ obtained after the last iteration ($m=45$). It can be found that $\Omega(t)$ and $\phi(t)$ are moderate in magnitude and continuous and smooth in shape, which confirms the experimental implementation [$\phi(t)$ can always be renormalized into the interval $[0,2\pi]$]. To investigate the influence of the deviation of $\Omega(t)$ and $\phi(t)$ on the cooling performance, we introduce the relative deviation $\zeta=(Q_{r}-Q_{t})/Q_{t}$, for $Q=\Omega$ or $\phi$, where $Q_{r}$ and $Q_{t}$ are, respectively, the realistic used parameters and the learned theoretical parameters. In Figs.~\ref{fig1}(d) and~\ref{fig1}(e), we show $\bar{N}_{\text{b}}(T)$ versus $\zeta$, and see that a slight deviation from $\zeta=0$ will cause a worse mechanical cooling for both the amplitude and phase. Meanwhile, the cooling performance has a weaker dependence on the phase deviation than the amplitude derivation. We analyze in Fig.~\ref{fig1}(f) the effect of the environment on the cooling performance after removing the pulsed drive at $T=38\omega_{\mathrm{m}}^{-1}$. The $\bar{N}_{\text{b}}$ will revert from $0.05$ to $20$ due to the action of the heat bath, and it will oscillate and approach equilibrium $\bar{N}_{\text{b}}\approx 9.7$ due to the optomechanical coupling. After the photons in the cavity are completely dissipated, the optomechanical interaction will no longer work, and the mean phonon number of the resonator will eventually be thermalized to $10^{3}$.

\begin{figure}[tbp]
\centering \includegraphics[width=0.49\textwidth]{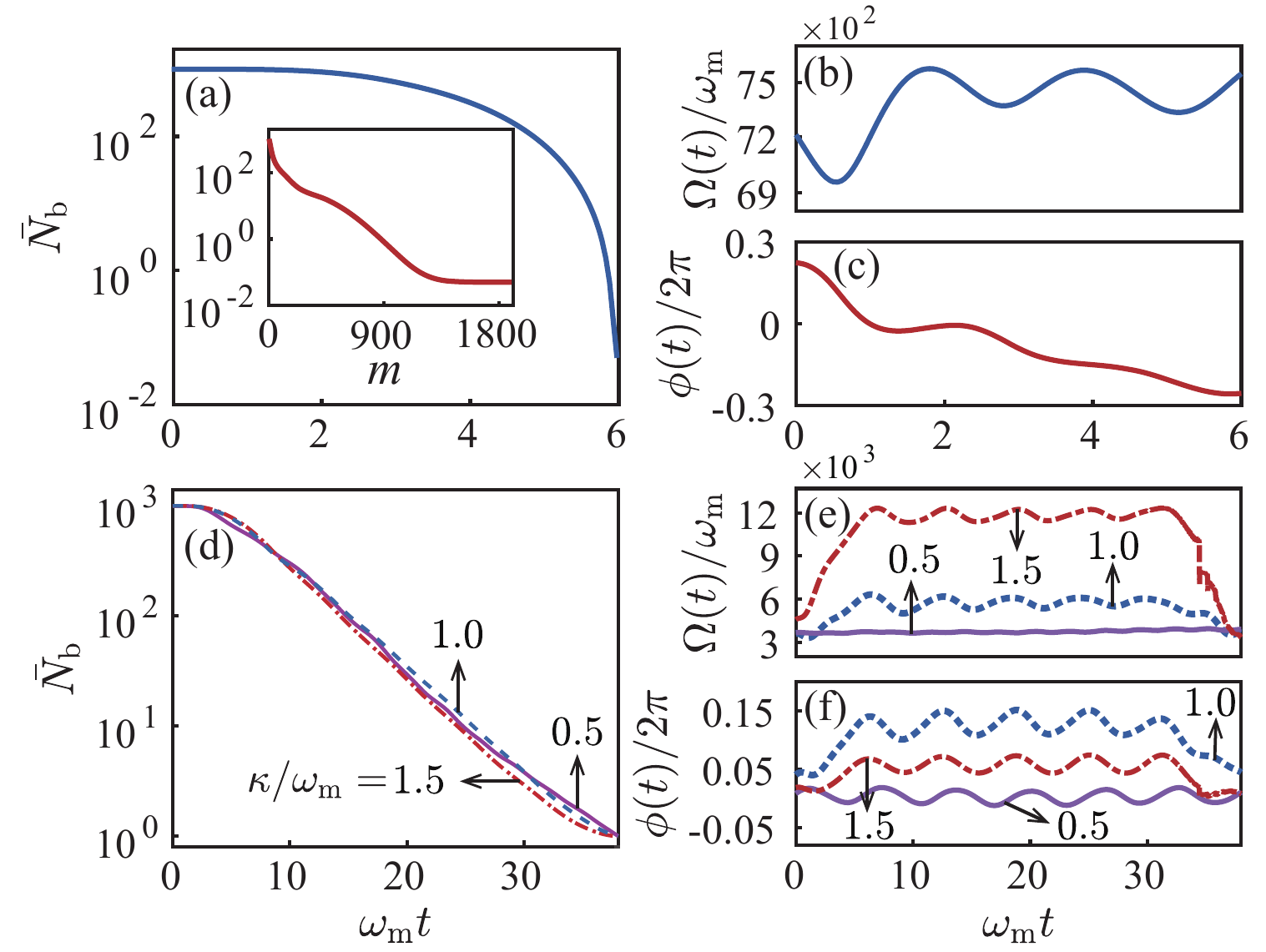}
\caption{(a) Mean phonon number $\bar{N}_{\text{b}}$ versus $\omega_{\mathrm{m}}t$ in one mechanical oscillation period ($T=6\omega_{\mathrm{m}}^{-1}$) for $m=1890$. The inset shows $\bar{N}_{\text{b}}(T=6\omega_{\mathrm{m}}^{-1})$ versus the iteration number $m$. (b) $\Omega(t)/\omega_{\mathrm{m}}$ and (c) $\phi(t)/2\pi$ versus $\omega_{\mathrm{m}}t$ obtained after the last iteration ($m=1890$) in (a). (d) $\bar{N}_{\text{b}}$ versus $\omega_{\mathrm{m}}t$ for different sideband-resolution parameters $\kappa/\omega_{\mathrm{m}}=$ 0.5, 1, and 1.5 when $\bar{N}_{\text{b}}(T=38\omega_{\mathrm{m}}^{-1})=1$. (e)~$\Omega(t)/\omega_{\mathrm{m}}$ and (f) $\phi(t)/2\pi$ vs $\omega_{\mathrm{m}}t$ corresponding to panel (d). Other parameters are the same as those in Fig.~\ref{fig1}.\label{fig2}}
\end{figure}
We also study the ultrafast cooling of the mechanical mode~\cite{TrianaPRL2016} by investigating the cooling within one mechanical oscillation period [Fig.~\ref{fig2}(a)]. Here we can see that the phonon occupation can be reduced from $10^{3}$ to $0.05$ within one oscillation period. The ultrafast cooling performance can also be improved by increasing the iteration number (see inset). In Figs.~\ref{fig2}(b) and~\ref{fig2}(c), we exhibit the waveforms of the driving amplitude and phase corresponding to Fig.~\ref{fig2}(a). We can see that a larger amplitude is required while the phase has no definite rules for the ground state cooling within one mechanical oscillation period. In addition, we study the dependence of cooling on the sideband-resolution condition. Figure~\ref{fig2}(d) shows that $\bar{N}_{\text{b}}$ consistently decreases from $10^{3}$ to $1$ for different sideband-resolution parameters. The
deterioration of the sideband-resolution parameter requires
a larger amplitude [Fig.~\ref{fig2}(e)] and has no clear phase feature
[Fig.~\ref{fig2}(f)]. The results indicate that the method works well in both the resolved-sideband and shallow unresolved-sideband regimes. Note that a large $\kappa/\omega_{\mathrm{m}}$ will dissipate the photons in the cavity within a short period of time, thus affecting the optomechanical interaction and causing the failure of ground state cooling.

\subsection{Large optomechanical entanglement}
\begin{figure}[tbp]
\centering \includegraphics[width=0.49\textwidth]{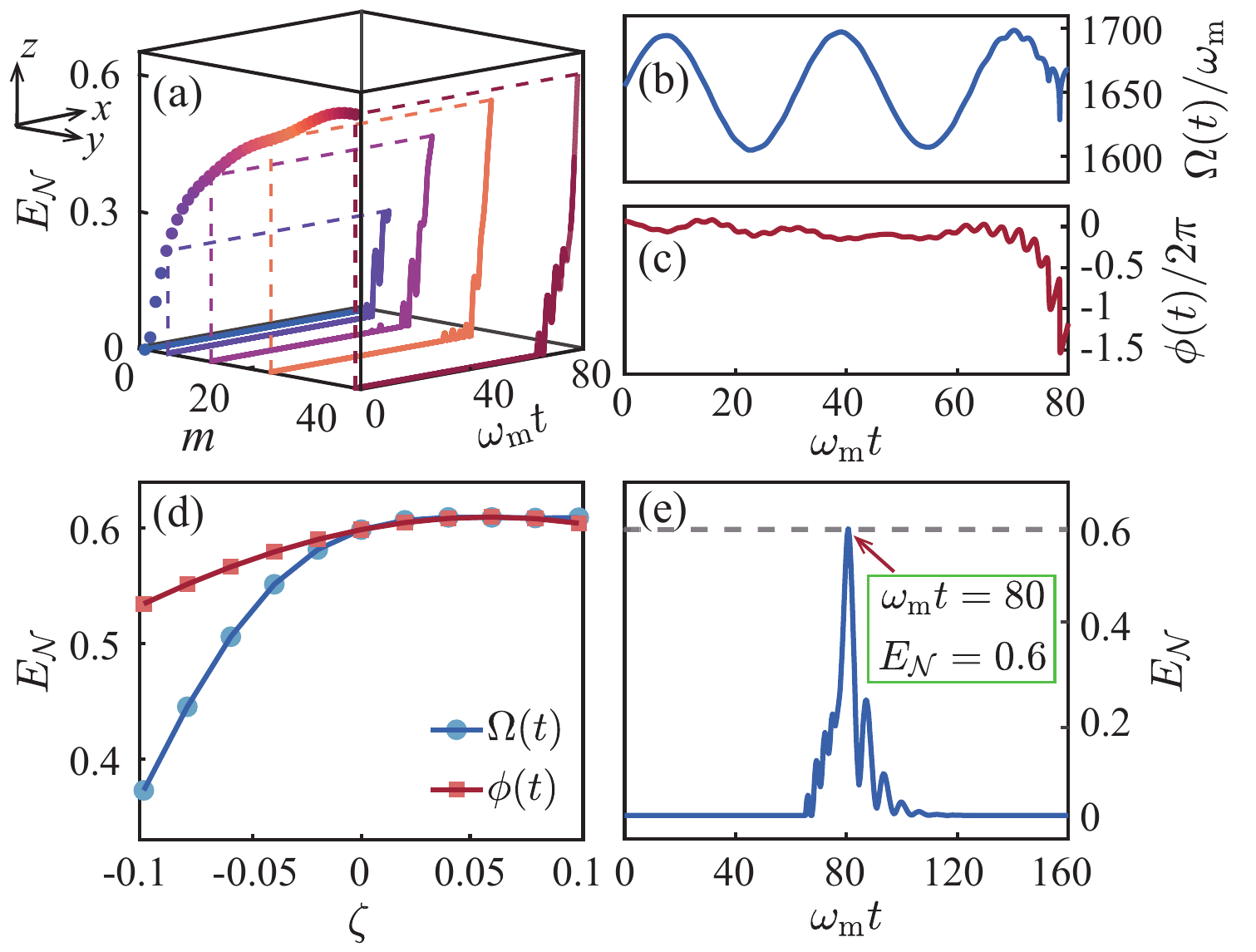}
\caption{(a) Logarithmic negativity $E_{\mathcal{N}}$ versus $\omega_{\mathrm{m}}t$ for different iteration numbers $m=0$, $5$, $13$, $24$, and $40$. The solid circles in the $yz$ plane show $E_{\mathcal{N}}(T)$ versus $m$. (b) $\Omega(t)/\omega_{\mathrm{m}}$ and (c) $\phi(t)/2\pi$ versus $\omega_{\mathrm{m}}t$ obtained after the last iteration ($m=40$). (d)~The influence of the driving amplitude and phase deviations on the entanglement performance at $T$. (e) Dynamic evolution of $E_{\mathcal{N}}$ after removing the pulsed drive. The gray dashed line marks the maximum entanglement value. Other parameters are $g_{0}/\omega_{\mathrm{m}}=4\times10^{-5}$, $\kappa/\omega_{\mathrm{m}}=0.2$, $\gamma_{\mathrm{m}}/\omega_{\mathrm{m}}=3\times10^{-6}$, $T=80\omega_{\mathrm{m}}^{-1}$, $\Delta_{\mathrm{c}}/\omega_{\mathrm{m}}=1$, and $\bar{n}_{\mathrm{m}}=100$. \label{fig3}}
\end{figure}

The optomechanical entanglement can be quantified by the logarithmic negativity~\cite{AdessoPRA2004,PlenioPRL205}
\begin{equation}
E_{\mathcal{N}}=\text{max}[0,-\ln(2\eta^{-})],
\end{equation}
where $\eta^{-}=\frac{1}{\sqrt{2}}[ \Sigma ( \mathbf{V}) -\sqrt{\Sigma (
\mathbf{V}) ^{2}-4\det \mathbf{V}}] ^{1/2}$ is the smallest
eigenvalue of the partial transpose of the covariance matrix $\mathbf{V}%
=[(\mathbf{V}_{A}, \mathbf{V}_{C});(\mathbf{V}_{C}^{\mathrm{T}}, \mathbf{V}_{B})]$, with $\Sigma ( \mathbf{V}) =\det
\mathbf{V}_{A}+\det \mathbf{V}_{B}-2\det \mathbf{V}_{C}$. A reduction in $\eta^{-}$ corresponds to an increase of the optomechanical entanglement. Therefore, the loss function is $\eta^{-}(T)$ for the entanglement problem, and the expression of the gradient-descent algorithm can be written as
\begin{equation}
\mathcal{Q}_{l+1}(s)=\mathcal{Q}_{l}(s)-\chi_{\mathcal{Q}}\frac{\delta\eta^{-}(T)}{\delta\mathcal{Q}_{l}(s)}.
\end{equation}%
Here, the variation of $\eta^{-}(T)$ with respect to $\mathcal{Q}%
(s)$ is given by
\begin{equation}  \label{eq13}
\small
\frac{\delta \eta ^{-}(T)}{\delta \mathcal{Q}(s)}\!=\!\frac{1}{4\eta ^{-}(T)%
\mathcal{Z}(T)}\left\{ \lambda (T) \frac{\delta \sum [\mathbf{V}(T)]}{\delta
\mathcal{Q}(s)}+\frac{2\delta \det \mathbf{V}(T)}{\delta \mathcal{Q}(s)}%
\right\} ,
\end{equation}
with
\begin{subequations}
\begin{align}
\mathcal{Z}(T)&=\sqrt{\Sigma [\mathbf{V}(T)]^{2}-4\det \mathbf{V}(T)}, \\
\lambda (T)&= \mathcal{Z}(T)-\sum [\mathbf{V}(T)].
\end{align}
\end{subequations}
Then, we need to calculate the values of $\delta \sum [\mathbf{V}(T)]/\delta
\mathcal{Q}(s)$ and $\delta [\det \mathbf{V}(T)]/\delta \mathcal{Q}(s)$,
which are essentially the variation of each covariance-matrix element with
respect to $\mathcal{Q}( s)$. As mentioned before, the term $\delta \mathbf{K}(s)/\delta \mathcal{Q}(s)$ in Eq.~(\ref{supeq321}) gives the variation of all covariance-matrix elements with respect to $%
\mathcal{Q}(s)$, and the value of $\delta  \eta ^{-}(T)/\delta \mathcal{Q}(s)$ can also be obtained. At this point, the gradient-descent algorithm can be executed until $\eta ^{-}(T)$ approaches a stable value. Note that the elements of $\mathbf{V}$ are transiently measurable in quadrature representation~\cite{Palomaki2013Science,LiuNPJquan,LHNp2019}, which allows us to experimentally obtain the transient optomechanical entanglement.

\begin{figure}[tbp]
\centering \includegraphics[width=0.49\textwidth]{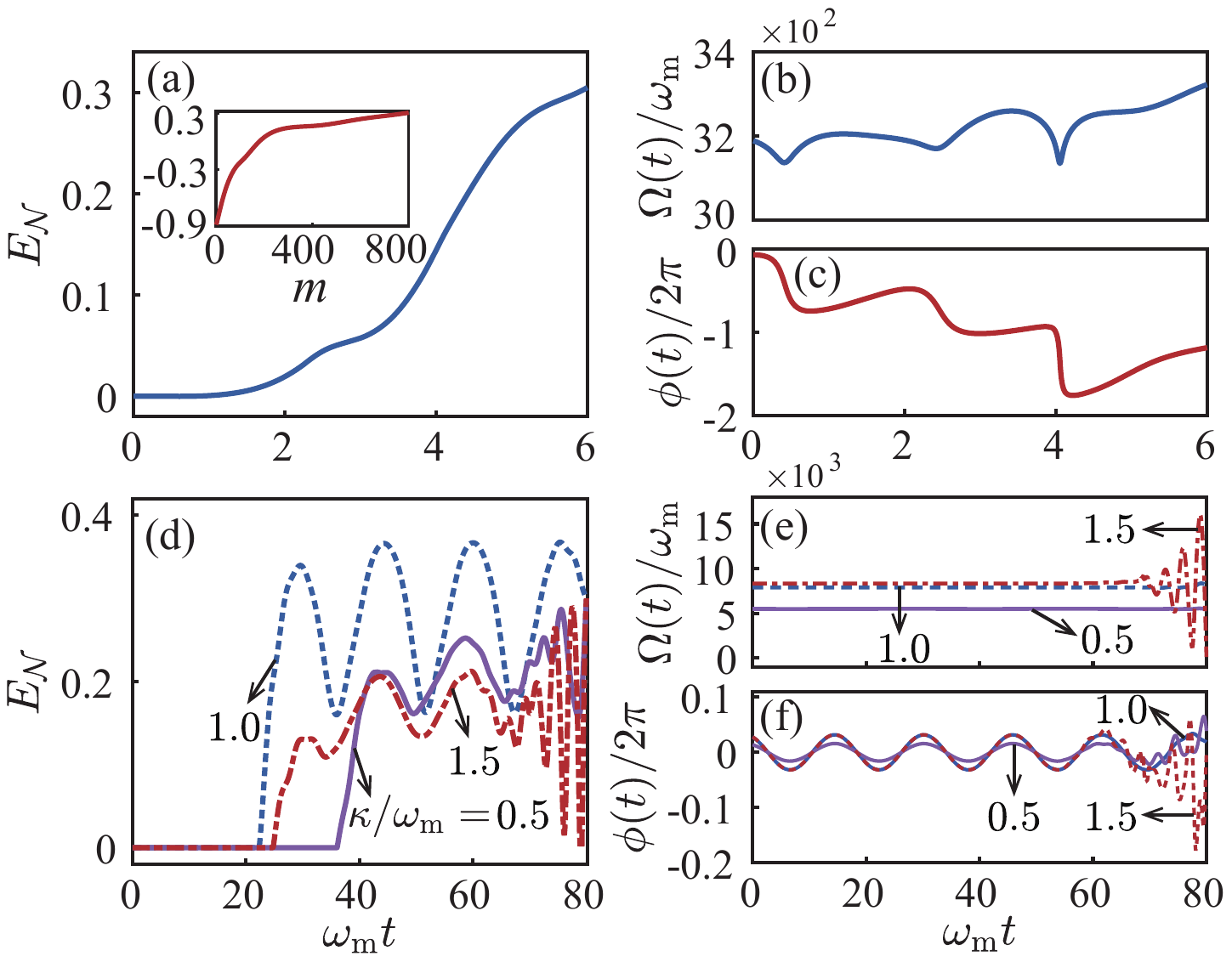}
\caption{(a) Logarithmic negativity $E_{\mathcal{N}}$ versus $\omega_{\mathrm{m}}t$ in one mechanical oscillation period ($T=6\omega_{\mathrm{m}}^{-1}$) under $\bar{n}_{\mathrm{m}}=10$ for $m=800$. The inset shows $E_{\mathcal{N}}(T=6\omega_{\mathrm{m}}^{-1})$ as a function of iteration number $m$. (b) $\Omega(t)/\omega_{\mathrm{m}}$ and (c) $\phi(t)/2\pi$ versus $\omega_{\mathrm{m}}t$ obtained after the last iteration ($m=800$) in (a). (d) $E_{\mathcal{N}}$ versus $\omega_{\mathrm{m}}t$ for different sideband-resolution parameters $\kappa/\omega_{\mathrm{m}}=$ 0.5, 1, and 1.5 under $E_{\mathcal{N}}(T=80\omega_{\mathrm{m}}^{-1})=0.3$. (e) Evolution of $\Omega(t)/\omega_{\mathrm{m}}$ and (f) $\phi(t)/2\pi$ corresponding to panel (d). Other parameters are the same as those in Fig.~\ref{fig3}.\label{fig4}}
\end{figure}
In Fig.~\ref{fig3}(a), we show the evolution of $E_{\mathcal{N}}$ for different iteration numbers. Here, $E_{\mathcal{N}}$ increases from a negligible value to $0.6$ as the iteration number increases.~In particular, the environmental thermal-excitation occupation of the mechanical resonator is $\bar{n}_{\mathrm{m}}=100$, indicating the thermal-noise resistance of the entanglement-preparation scheme.~The waveforms of the pulse drive corresponding to $E_{\mathcal{N}}=0.6$ is shown in~Figs.~\ref{fig3}(b) and~\ref{fig3}(c). The entanglement $E_{\mathcal{N}}=0.6$ is about two to four times over the previously reported steady-state entanglement~\cite{Vitali2007PRL,Genes2008PRA,Palomaki2013Science,Jiao2020PRL,Lai2022PRL} (about 0.15 to 0.3) with experimentally accessible drives.~Note that a larger optomechanical entanglement can be achieved by varying the iteration parameters, and the numerical results show that this requires a larger driving amplitude [$E_{\mathcal{N}}=0.9$ requires $|\Omega_{\text{max}}|/\omega_{\mathrm{m}}\approx4330$; see Sec.~\ref{sec4}(C) for details].

The influence of the driving parameter deviation of the drive on the entanglement performance is shown in Fig.~\ref{fig3}(d). Here, we can see that the dependences of $E_{\mathcal{N}}$ on both the amplitude and phase show similar trends. For the relative deviation $\zeta\in[-0.1,0.1]$, $E_{\mathcal{N}}$ is an increasing function of $\zeta$. In particular, the rate of change in the lower deviation region ($\zeta\!<\!0$) are much steeper than that in the upper deviation~($\zeta\!>\!0$). Figure~\ref{fig3}(e) shows the action of the environment on entanglement after removing the pulsed drive. The $E_{\mathcal{N}}$ goes from 0.6 to 0 very fast ($30\omega_{\mathrm{m}}^{-1}$) because the mechanical resonator is heated.

An ultrafast entanglement generation ($E_{\mathcal{N}}=0.3$) within one mechanical oscillation period can also be realized with the optimization approach [Fig.~\ref{fig4}(a)]. The entanglement can also be improved by increasing the iteration number (see inset). Note that here the $E_{\mathcal{N}}$ is almost zero when $m\leq180$. In Figs.~\ref{fig4}(b) and~\ref{fig4}(c), we show the waveforms of the driving amplitude and phase corresponding to the dynamics of $E_{\mathcal{N}}$ in Fig.~\ref{fig4}(a), which are moderate in size and continuous in shape. We also show the evolution of $E_{\mathcal{N}}$ for different sideband-resolution parameters in Fig.~\ref{fig4}(d), and the deterioration of sideband-resolution parameter requires a large driving amplitude to achieve $E_{\mathcal{N}}=0.3$ [Fig.~\ref{fig4}(e)], while the phase dependence has no specific feature [Fig.~\ref{fig4}(f)]. The results indicate that the quantum-learning-control method is valid in generating optomechanical entanglement in both resolved-sideband and shallow unresolved-sideband regimes.

\section{Discussions\label{sec4}}
In this section, we present some discussions on the dynamics of both the displacement amplitude and normalized detuning, impact of laser amplitude and phase noises on the cooling and entanglement performances, deeper optomechanical cooling and larger optomechanical entanglement, and
experimental implementation of our scheme.

\begin{figure*}[tbp]
\centering \includegraphics[width=1\textwidth]{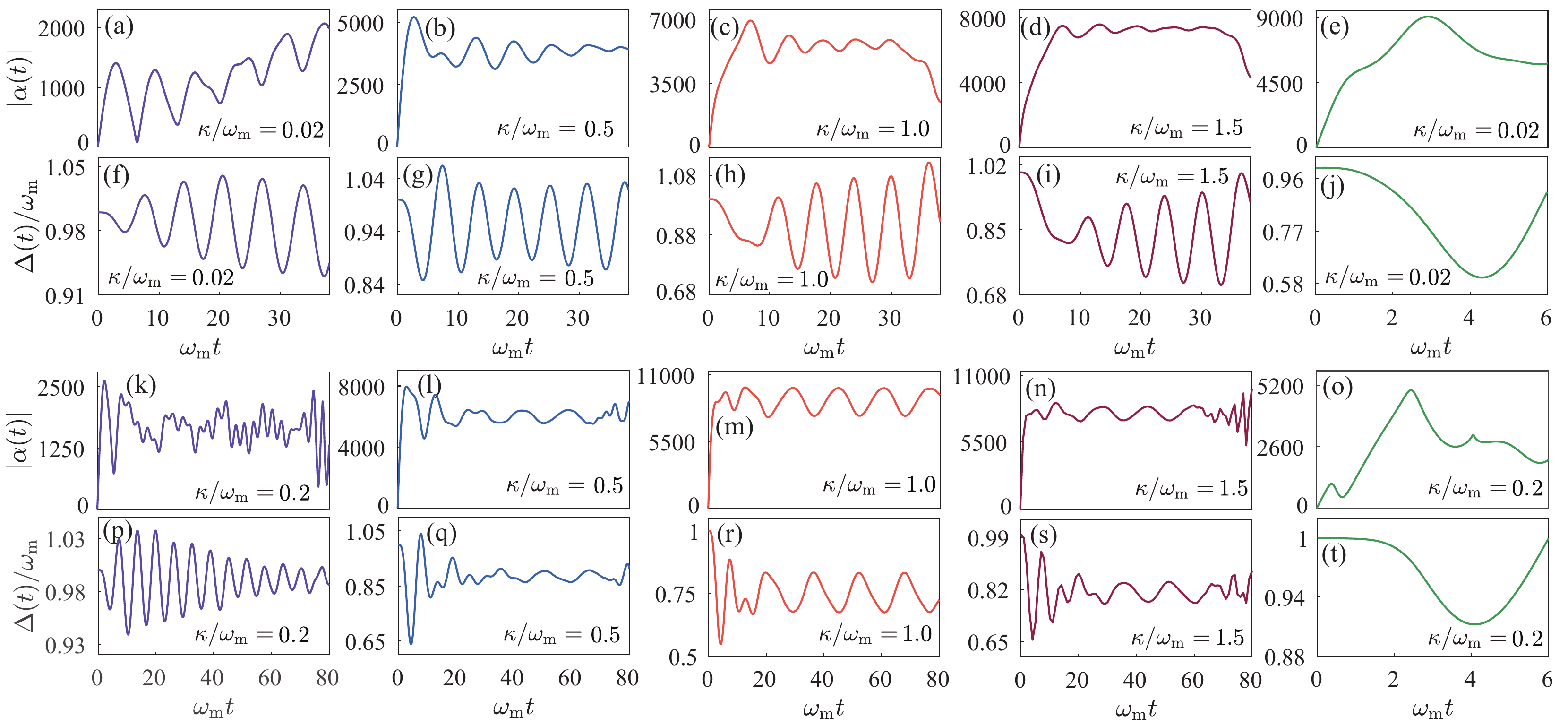}
\caption{(a)-(d) Modulus of the cavity-field displacement amplitude $|\protect%
\alpha(t)|$ and (f)-(i) normalized detuning $\Delta(t)/\protect\omega_{\mathrm{%
m}}$ versus the scaled evolution time $\protect\omega_{\mathrm{m}}t$ for
optomechanical cooling at different sideband-resolution parameters $\protect%
\kappa/\protect\omega_{\mathrm{m}}=0.02, 0.5, 1.0$, and $1.5$, for $T=38%
\protect\omega_{\mathrm{m}}^{-1}$. (e) $|\protect\alpha(t)|$ and (j) $%
\Delta(t)/\protect\omega_{\mathrm{m}}$ as a function of $\protect\omega_{\mathrm{m}}t$
during one mechanical oscillation period ($T=6\protect\omega_{\mathrm{m}%
}^{-1}$), for $\protect\kappa/\protect\omega_{\mathrm{m}}=0.02$. Other
parameters are $g_{0}/\protect\omega_{\mathrm{m}}=4\times10^{-5}$, $\protect%
\gamma_{\mathrm{m}}/\protect\omega_{\mathrm{m}}=3\times10^{-6}$, $\Delta_{%
\mathrm{c}}/\protect\omega_{\mathrm{m}}=1$, and $\bar{n}_{\mathrm{m}}=10^{3}$%
. (k)-(n) Modulus of the cavity-field displacement amplitude $|\protect%
\alpha(t)|$ and (p)-(s) normalized detuning $\Delta(t)/\protect\omega_{\mathrm{%
m}}$ versus the scaled evolution time $\protect\omega_{\mathrm{m}}t$ for
optomechanical entanglement at different sideband-resolution parameters $%
\protect\kappa/\protect\omega_{\mathrm{m}}=0.2, 0.5, 1.0$, and $1.5$, for $%
T=80\protect\omega_{\mathrm{m}}^{-1}$. (o) $|\protect\alpha(t)|$ and (t) $%
\Delta(t)/\protect\omega_{\mathrm{m}}$ as a function of $\protect\omega_{\mathrm{m}}t$
during one mechanical oscillation period ($T=6\protect\omega_{\mathrm{m}%
}^{-1}$), for $\protect\kappa/\protect\omega_{\mathrm{m}}=0.2$ and $\bar{n}%
_{\mathrm{m}}=10$. Other parameters are $g_{0}/\protect\omega_{\mathrm{m}%
}=4\times10^{-5}$, $\protect\kappa/\protect\omega_{\mathrm{m}}=0.2$, $%
\protect\gamma_{\mathrm{m}}/\protect\omega_{\mathrm{m}}=3\times10^{-6}$, $%
\Delta_{\mathrm{c}}/\protect\omega_{\mathrm{m}}=1$, and $\bar{n}_{\mathrm{m}%
}=100$.}
\label{SubFig1}
\end{figure*}

\subsection{Dynamics of both the displacement amplitude and the normalized detuning}
Unlike the sideband-cooling case, the present displacement amplitude $\alpha(t)$ and normalized detuning $\Delta(t)$ of the cavity field take transient values. To verify the linearization condition and the red-sideband resonance
condition, we plot in Fig.~\ref{SubFig1} the dynamic
evolution of the displacement amplitude $\alpha (t)$ and the normalized
detuning $\Delta (t)$ for optomechanical cooling and optomechanical
entanglement. As shown in Figs.~\ref{SubFig1}(a)-(e) and~\ref{SubFig1}(k)-(o), the modulus of the cavity-field displacement amplitude $%
|\alpha (t)|$ for optomechanical cooling and entanglement is much larger than the one in the time domain. Thus, the linearization conditions are satisfied in both the
cooling and entanglement schemes for our parameters. In particular, the
system enters the strong-coupling regime, $|G|>\{\kappa ,\gamma _{\mathrm{m}}\}\rightarrow |\alpha |>\mathrm{max}\{\{\kappa ,\gamma _{\mathrm{m}}\}/g_{0}\}=500$,
in most of the duration. Additionally, the normalized detuning $\Delta (t)/\omega _{\mathrm{m}}$
oscillates gradually and moves away from $1$ as the sideband-resolution
condition is not satisfied [Figs.~\ref{SubFig1}(f) to~\ref{SubFig1}(i) and~\ref{SubFig1}(p) to~\ref{SubFig1}(s)]. Similarly, the $\Delta (t)/\omega _{\mathrm{m}}$ corresponding to optomechanical
cooling and entanglement within one mechanical oscillation period also
deviates from 1 [Figs.~\ref{SubFig1}(j) and~\ref{SubFig1}(t)].
\begin{figure}[tbp]
\centering \includegraphics[width=0.48\textwidth]{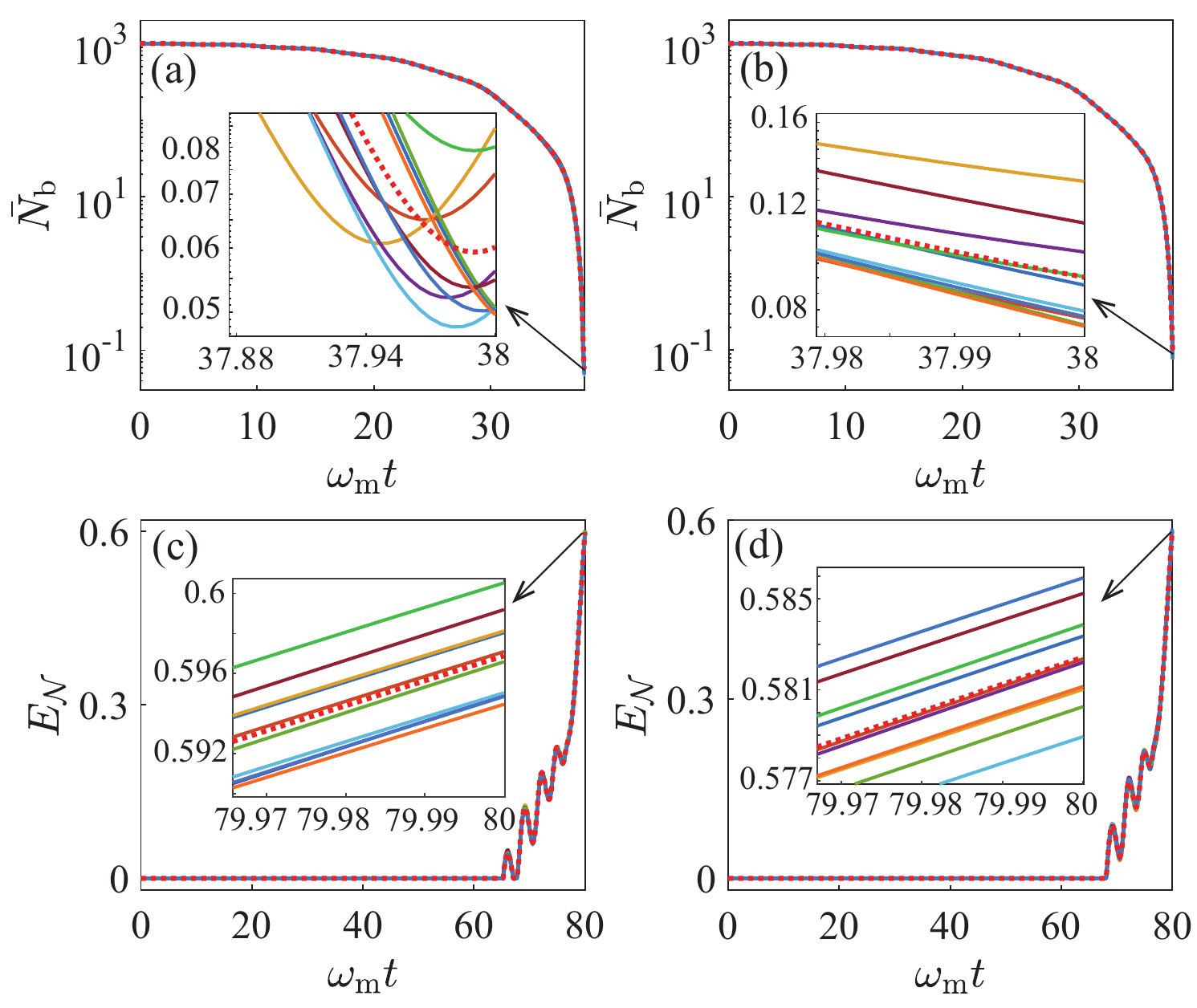}
\caption{The mean phonon number $\bar{N}_{\text{b}}$ as a function of the scaled evolution time $\omega_{\mathrm{m}}t$ after adding (a) laser amplitude and (b) phase noises. The driving amplitude is added to the random number of a normal distribution with an average value of 0 and a standard deviation of 200; while the driving phase is added to the random number of a normal distribution with an average value of 0 and a standard deviation of 0.1. Other parameters used are the same as in Fig.~\ref{fig1}. The logarithmic negativity $E_{\mathcal{N}}$ versus $\omega_{\mathrm{m}}t$ after adding (c) laser amplitude and (d) phase noise. The driving amplitude is incorporated to the random number of a normal distribution with an average value of 0 and a standard deviation of 200; while the driving phase is added to the random number of a normal distribution with an average value of 0 and a standard deviation of 0.2. Other parameters used are the same as in Fig.~\ref{fig3}.}
\label{SubFig2}
\end{figure}
\begin{figure*}[htbp]
\centering \includegraphics[width=1\textwidth]{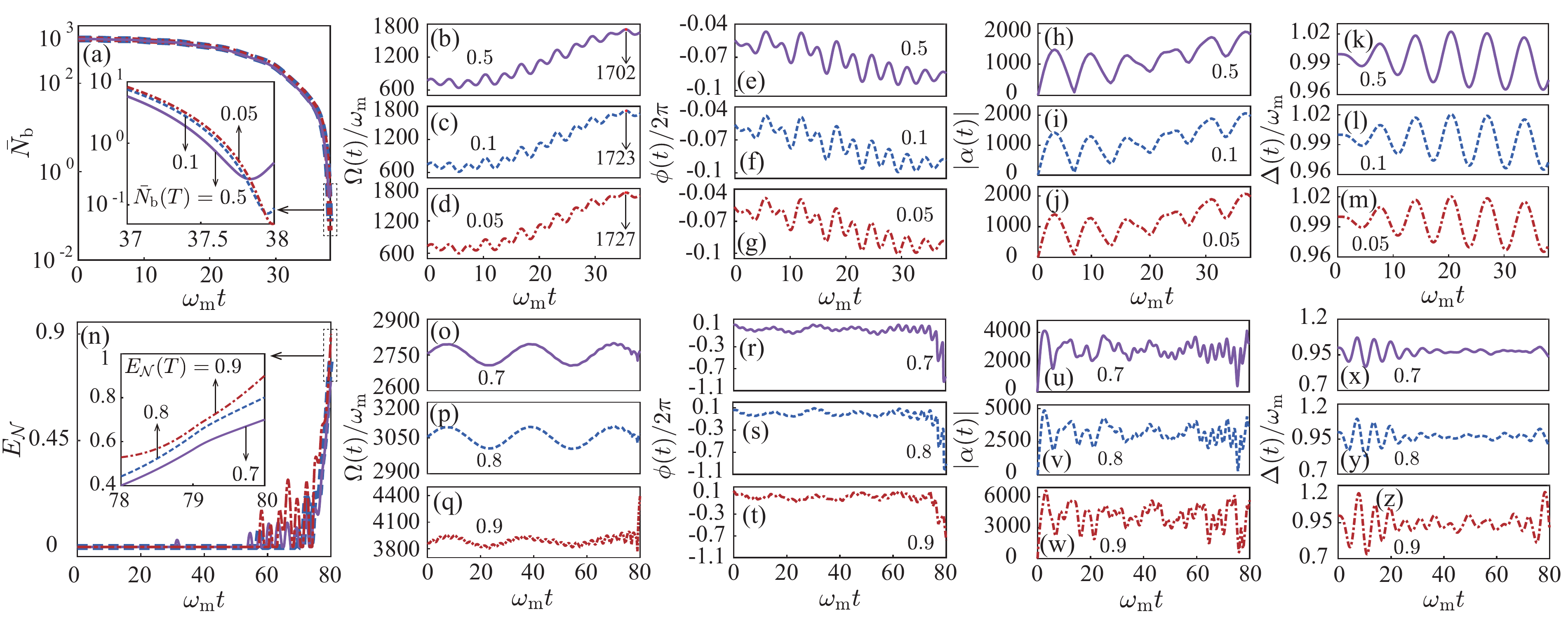}
\caption{(a) Dynamic evolution of the mean phonon number $\bar{N}_{\text{b}}$ under different cooling targets. (b)-(d) The normalized driving amplitude $\Omega(t)/\omega_{\mathrm{m}}$ and (e)-(g) driving phase $\phi(t)/2\pi$ evolve with the scaling time $\omega_{\mathrm{m}}t$ corresponding to different cooling targets in panel (a). (h)-(j) The modulus of the cavity-field displacement amplitude $|\alpha(t)|$ and (k)-(m) the normalized detuning $\Delta(t)/\omega_{\mathrm{m}}$ versus the scaled evolution time $\omega_{\mathrm{m}}t$ corresponding to panel (a). Here, the parameters are $g_{0}/\omega_{\mathrm{m}}=4\times10^{-5}$, $\kappa/\omega_{\mathrm{m}}=0.02$, $\gamma_{\mathrm{m}}/\omega_{\mathrm{m}}=3\times10^{-6}$, $\Delta_{\mathrm{c}}/\omega_{\mathrm{m}}=1$, $T=38\omega_{\mathrm{m}}^{-1}$, and $\bar{n}_{\mathrm{m}}=10^{3}$. (n) Dynamic evolution of the logarithmic negativity $E_{\mathcal{N}}$  with different given entanglement degrees. (o)-(q) $\Omega(t)/\omega_{\mathrm{m}}$ and (r)-(t) $\phi(t)/2\pi$ versus $\omega_{\mathrm{m}}t$ for different entanglement degrees in panel (n). (u)-(w) $|\alpha(t)|$ and (x)-(z) $\Delta(t)/\omega_{\mathrm{m}}$ versus $\omega_{\mathrm{m}}t$ corresponding to panel (n). The parameters used are $g_{0}/\omega_{\mathrm{m}}=4\times10^{-5}$, $\kappa/\omega_{\mathrm{m}}=0.2$, $\gamma_{\mathrm{m}}/\omega_{\mathrm{m}}=3\times10^{-6}$, $\Delta_{\mathrm{c}}/\omega_{\mathrm{m}}=1$, $T=80\omega_{\mathrm{m}}^{-1}$, and $\bar{n}_{\mathrm{m}}=100$.\label{SubFig3}}
\end{figure*}
\subsection{Impact of laser amplitude and phase noises on the cooling and entanglement performances}
In realistic cases, the classical driving lasers inevitably have phase and amplitude noises. In particular, numerous studies demonstrated that phase noise has destructive effects on quantum phenomena~\cite{MeenehanPRX2015,ShomroniPRL,Zeng2022,Meyer2019}. To assess the influence of laser amplitude and phase noises on the cooling and entanglement performance, we evaluate the influence of laser amplitude and phase noises by adding Gaussian random numbers (white noise) to each time of the resulting pulse drive amplitude and phase.

In Figs.~\ref{SubFig2}(a) and~\ref{SubFig2}(b), we show the mean phonon number $\bar{N}_{\text{b}}$ versus the scaled evolution time $\omega_{\mathrm{m}}t$ after adding Gaussian random numbers to the driving amplitude and phase, respectively. The ten solid lines represent the evolution of $\bar{N}_{\text{b}}$ under ten distinct random conditions, while the red dotted line represents the average of these ten evolutions. It is evident that the presence of random laser noise minimally impacts the cooling results, proving that laser noise affects the cooling performance but not significantly. Similarly, we display in Figs.~\ref{SubFig2}(c) and~\ref{SubFig2}(d) the dynamic evolution the logarithmic negativity $E_{\mathcal{N}}$ with the incorporation of Gaussian random numbers into the driving amplitude and phase, respectively. These results reveal the robustness of optomechanical entanglement with respect to random laser amplitude and phase noises. Note that the laser phase noise acts on the mechanical resonator as an additional heating noise proportional to the intracavity field amplitude, which means that we can consider the optical heating effect together with the phase noise~\cite{LPRA2008}.

\subsection{Deeper optomechanical cooling and larger optomechanical entanglement}
In this subsection, we discuss the conditions for achieving deeper
optomechanical cooling and larger optomechanical entanglement. Concretely,
we present the dynamic evolution of both the mean phonon number $\bar{N}_{%
\text{b}}$ and the logarithmic
negativity $E_{\mathcal{N}}$ under different given targets.~We
also show the waveforms of the driving amplitude and phase corresponding to
these cases.

In Fig.~\ref{SubFig3}(a), we plot the $\bar{N}_{\text{b}}$ as a function of
the normalized evolution time $\omega_{\mathrm{m}}t$ for different given cooling
targets. It can be seen from the inset in Fig.~\ref{SubFig3}(a) that
different cooling targets [$\bar{N}_{\text{b}}(T)=0.5$, 0.1, 0.05] are
achieved at the target time $T=38\omega_{\mathrm{m}}^{-1}$, which proves
that the deeper ground-state cooling of the mechanical resonator can be
realized using the quantum-learning-control method. The corresponding
waveforms of the driving amplitude and phase used to realize the cooling
target in Fig.~\ref{SubFig3}(a) are shown in Figs.~\ref{SubFig3}(b) to~\ref{SubFig3}(d) and~%
\ref{SubFig3}(e) to~\ref{SubFig3}(g). It can be seen that, for a deeper cooling of the
mechanical resonator, a larger driving amplitude is needed [Figs.~\ref{SubFig3}(b) to~\ref{SubFig3}(d)] while the phase varies irregularly~[Figs.~\ref{SubFig3}(e) to~\ref{SubFig3}(g)]. We can understand this point from the perspective of energy. For the
same target time $T$, it needs much energy to achieve a better cooling
target. Since the driving amplitude and driving
power of the pulse field satisfy the relationship $|\Omega|=\sqrt{2\mathbf{P}%
\kappa/\hbar\omega_{\mathrm{L}}}$ ($\mathbf{P}$ is the driving power of the
pulse field), then the better the cooling target, the
larger the pulsed field amplitude.

To verify the linearization condition and the red-sideband resonance
condition, we numerically simulate the dynamic evolution of both the
displacement amplitude $\alpha(t)$ and the normalized detuning $\Delta(t)$
corresponding to the pulsed drive in Figs.~\ref{SubFig3}(b) to~\ref{SubFig3}(d) and~\ref{SubFig3}(e) to~\ref{SubFig3}(g). At this time, the displacement amplitude $|\alpha(t)|$ of the
cavity field can be considered to be much larger than one in the time domain [Figs.~\ref{SubFig3}(h) to~\ref{SubFig3}(j)], and the
normalized detuning $\Delta(t)/\omega_{\mathrm{m}}$ oscillates around $1$
[Figs.~\ref{SubFig3}(k) to~\ref{SubFig3}(m)]. Hence, both the linearization condition and the
red-sideband resonance condition are satisfied approximately for the
parameters used here.

Figure~\ref{SubFig3}(n) shows the $E_{\mathcal{N}}$ as a function of the
normalized evolution time $\omega_{\mathrm{m}}t$ for different given
entanglement degrees. We can see that different degrees of optomechanical
entanglement [$E_{\mathcal{N}}(T)=0.7, 0.8$, and 0.9] can be achieved at the
target time $T=80\omega_{\mathrm{m}}^{-1}$ through the
quantum-learning-control method. We also present the waveforms of the driving amplitude and phase used to
generate the optomechanical entanglement in Fig.~\ref{SubFig3}(n), as shown
in Figs.~\ref{SubFig3}(o) to~\ref{SubFig3}(q) and~\ref{SubFig3}(r) to~\ref{SubFig3}(t). It can be seen that the
increase of the optomechanical entanglement requires an increasing driving
amplitude, while the phase has small change. This is because the
entanglement depends strongly on the driving amplitude with the enhancement
of the entanglement degree, so the phase correction is small. In particular,
the optomechanical entanglement can reach $E_{\mathcal{N}}=0.9$ when the
driving amplitude reaches $|\Omega_{\text{max}}|/\omega_{\mathrm{m}%
}\approx4330$. Similar to the cooling case, we can also explain why the increase of optomechanical entanglement requires a
large driving amplitude from the viewpoint of
energy. In addition, we show the modulus of the displacement
amplitude $|\alpha(t)|$ and the normalized detuning $\Delta(t)/\omega_{%
\mathrm{m}}$ as functions of $\omega_{\mathrm{m}}t$, corresponding to the
entanglement degrees in Fig.~\ref{SubFig3}(n). In this case, both the
linearization condition [Figs.~\ref{SubFig3}(u) to~\ref{SubFig3}(w)] and the red-sideband
resonance condition are approximately satisfied [Figs.~\ref{SubFig3}(x) to~\ref{SubFig3}(z)] as
the entanglement increases.

\subsection{Experimental implementation}
Finally we present some discussions on the experimental implementation of our scheme. The optimization method works for a general LGQ system and hence it can be widely used in various linear physical setups. In particular, our proposed optimized optomechanical cooling and entanglement methods can be demonstrated in various experimental systems~\cite{Kippenberg2014RMP}, such as photonic crystal nanobeams~\cite{Chan2011Nature,Riedinger2016Nature,Eichenfield2009,ChanNT2009}, electromechanical systems~\cite{Teufel2011Nature,Palomaki2013Science,HertzbergNP2010,MasselN2011,MasselNC2012}, optical microresonators~\cite{ParkNP2009,SchliesserNP2009,VerhagenNature2012}, and  Fabry-P\'{e}rot cavities~\cite{GiganN2006,ArcizetN2006,KlecknerN2006,HammererNature2009}. Currently, the linearized optomechanical systems have been implemented in many experimental setups. To control the optomechanical system, pulsed fields are usually introduced to drive the optomechanical cavity. The suggested drive pulses used in our scheme are continuous and smooth, and hence are experimentally accessible. The above analyses indicate that our scheme should be within the reach of the current experimental conditions.

\section{Conclusion\label{sec5}}
In conclusion, we proposed a general optimal control method for LGQ systems based on the gradient-descent algorithm. This approach has been successfully applied to cavity optomechanical systems, demonstrating the realization of both deep optomechanical cooling and large optomechanical entanglement. The
optomechanical cooling can exceed the sideband-cooling limit in the continuous-wave-driven strong-coupling case, and the optomechanical entanglement can surpass several times the corresponding steady-state entanglement under dozens of thermal phonons. We also obtained the optimal driving amplitude and phase. This work opens an avenue for quantum learning control of LGQ systems, and possesses a wide application in modern quantum science.

\begin{acknowledgments}
J.-Q.L. was supported, in part, by the National Natural Science Foundation of China (Grants No.~12175061, No.~12247105, and No.~11935006), the Science and Technology Innovation Program of Hunan Province (Grant No.~2021RC4029), and the Hunan Provincial Major Science and Technology Program (Grant No.~2023ZJ1010). F. N. is supported, in part, by Nippon Telegraph and Telephone Corporation (NTT) Research, the Japan Science and Technology Agency (JST) via the Quantum Leap Flagship Program (Q-LEAP) and the Moonshot R$\&$D under Grant No.~JPMJMS2061, the Asian Office of Aerospace Research and Development (AOARD) under Grant No.~FA2386-20-1-4069, and the Office of Naval Research (ONR) Grant No. N62909-23-1-2074. D.D. acknowledges the support of the Australian Research Council's Future Fellowship funding scheme under Project No.~FT220100656 and the U.S. Office of Naval Research Global under Grant No.~N62909-19-1-2129. Q.S.T. acknowledges support from NSFC (Grant No.~12275077).
\end{acknowledgments}


\begin{thebibliography}{99}
\bibitem{WisemanPRL2005} H. M. Wiseman and A. C. Doherty, Optimal Unravellings for Feedback Control in Linear Quantum Systems, \href{https://doi.org/10.1103/PhysRevLett.94.070405}{Phys. Rev. Lett. \textbf{94}, 070405 (2005)}.

\bibitem{Wiseman2009}H. M. Wiseman and G. J. Milburn, \emph{Quantum Measurement and Control} (Cambridge University Press, Cambridge, England, 2009).

\bibitem{Nurdin2017} H. Nurdin and N. Yamamoto, \emph{Linear Dynamical Quantum Systems: Analysis, Synthesis, and Control}, Communications and Control Engineering (Springer, NewYork, 2017).

\bibitem{Zhang2022} G. Zhang and Z. Dong, Linear quantum systems: A tutorial, \href{https://doi.org/10.1016/j.arcontrol.2022.04.013}{Annu. Rev. Control \textbf{54}, 274 (2022)}.

\bibitem{WeedbrookRMP2012} C. Weedbrook, S. Pirandola, R. Garc\'{\i}a-Patr\'{o}n, N. J. Cerf, T. C.
Ralph, J. H. Shapiro, and S. Lloyd, Gaussian quantum information, \href{https://doi.org/10.1103/RevModPhys.84.621}{Rev. Mod. Phys. \textbf{84}, 621 (2012)}.

\bibitem{Peres2002} A. Peres, \emph{Quantum Theory: Concepts and Methods} (Springer, Dordrecht, The Netherlands, 2002).

\bibitem{Alessandro2007} D. d'Alessandro, \emph{Introduction to Quantum Control and Dynamics} (CRC Press, Boca Raton, FL, 2007).

\bibitem{Braunstein2005} S. Braunstein and P. van Loock, Quantum information with continuous variables, \href{https://doi.org/10.1103/RevModPhys.77.513}{Rev. Mod. Phys. \textbf{77}, 513 (2005)}.

\bibitem{Wangpr2007} X. B. Wang, T. Hiroshimab, A. Tomitab, and M. Hayashi, Quantum information with Gaussian states, \href{https://doi.org/10.1016/j.physrep.2007.04.005}{Phys. Rep. \textbf{448}, 1 (2007)}.

\bibitem{Schwab2012PT} M. Aspelmeyer, P. Meystre, and K. Schwab, Quantum
optomechanics, \href{https://doi.org/10.1063/PT.3.1640}{Phys. Today \textbf{65}(7), 29 (2012)}.

\bibitem{Kippenberg2014RMP} M. Aspelmeyer, T. J. Kippenberg, and F.
Marquardt, Cavity optomechanics, \href{https://doi.org/10.1103/RevModPhys.86.1391}{Rev. Mod. Phys. \textbf{86}, 1391 (2014)}.

\bibitem{Bowen2016} W. P. Bowen and G. J. Milburn, \emph{Quantum Optomechanics} (CRC Press, Boca Raton, FL, 2016).

\bibitem{Mandel1965} L. Mandel and E. Wolf, Coherence properties of optical fields, \href{https://doi.org/10.1103/RevModPhys.37.231}{Rev. Mod. Phys. \textbf{37}, 231 (1965)}.

\bibitem{Lukin2003} M. D. Lukin, Colloquium: Trapping and manipulating photon states in atomic ensembles, \href{https://doi.org/10.1103/RevModPhys.75.457}{Rev. Mod. Phys. \textbf{75}, 457 (2003)}.

\bibitem{HammererRMP2010} K. Hammerer, A. S. S{\o}rensen, and E. S. Polzik, Quantum interface between light and atomic ensembles, \href{https://doi.org/10.1103/RevModPhys.82.1041}{Rev. Mod. Phys. \textbf{82}, 1041 (2010)}.

 \bibitem{Dalfovo1999} F. Dalfovo, S. Giorgini, L. P. Pitaevskii, and S. Stringari, Theory of Bose-Einstein condensation in trapped gases, \href{https://doi.org/10.1103/RevModPhys.71.463}{Rev. Mod. Phys. \textbf{71}, 463 (1999)}.

\bibitem{Morsch2006} O. Morsch and M. Oberthaler, Dynamics of Bose-Einstein condensates in optical lattices, \href{https://doi.org/10.1103/RevModPhys.78.179}{Rev. Mod. Phys. \textbf{78}, 179 (2006)}.

\bibitem{BlochNJP2018} A. Ac\'{\i}n, I. Bloch, H. Buhrman, T. Calarco, C. Eichler, J.
Eisert, D. Esteve, N. Gisin, S. J. Glaser, F. Jelezko, S. Kuhr,
M. Lewenstein, M. F. Riedel, P. O. Schmidt, R. Thew, A.
Wallraff, I. Walmsley, and F. K. Wilhelm, The quantum technologies roadmap: A European community view, \href{https://iopscience.iop.org/article/10.1088/1367-2630/aad1ea}{New J. Phys. \textbf{20}, 080201 (2018)}.

\bibitem{Benyoucef2023}  M. Benyoucef, \emph{Photonic Quantum Technologies: Science and Applications} (Wiley, Weinheim, Germany, 2023).

 \bibitem{Kalman1960}  R. E. Kalman, A new approach to linear filtering and prediction problems, \href{https://doi.org/10.1115/1.3662552}{J. Basic Eng. \textbf{82}, 35 (1960)}.

 \bibitem{Nurdin2009} H. Nurdin, M. R. James, and I. R. Petersen, Coherent quantum LQG control, \href{https://doi.org/10.1016/j.automatica.2009.04.018}{Automatica \textbf{45}, 1837 (2009)}.

 \bibitem{Dong2010}   D. Dong and I. R. Petersen, Quantum control theory and applications: A survey, \href{https://digital-library.theiet.org/content/journals/10.1049/iet-cta.2009.0508}{IET Control Theor. Appl. \textbf{4}, 2651 (2010)}.

 \bibitem{ZhangRP2017} J. Zhang, Y.-X. Liu, R.-B. Wu, K. Jacobs, and F. Nori, Quantum feedback: Theory, experiments, and applications, \href{https://doi.org/10.1016/j.physrep.2017.02.003}{Phys. Rep. \textbf{679}, 1 (2017)}.

\bibitem{ZhangIEEETrans2011} G. Zhang and M. R. James, Direct and Indirect Couplings in Coherent Feedback Control of Linear Quantum Systems, \href{https://doi.org/10.1109/TAC.2010.2096010}{IEEE Trans. Autom. Control \textbf{56}, 1535 (2011)}.

\bibitem{Yamamotoprx2014} N. Yamamoto, Coherent versus Measurement Feedback: Linear Systems Theory for Quantum Information, \href{https://doi.org/10.1103/PhysRevX.4.041029}{Phys. Rev. X \textbf{4}, 041029 (2014)}.

\bibitem{WieczorekPRL2015} W. Wieczorek, S. G. Hofer, J. Hoelscher-Obermaier, R. Riedinger, K. Hammerer, and M. Aspelmeyer, Optimal State Estimation for Cavity Optomechanical Systems, \href{https://doi.org/10.1103/PhysRevLett.114.223601}{Phys. Rev. Lett. \textbf{114}, 223601 (2015)}.

 \bibitem{ManciniPRA2007} S. Mancini and H. M. Wiseman, Optimal control of entanglement via quantum feedback, \href{https://doi.org/10.1103/PhysRevA.75.012330}{Phys. Rev. A \textbf{75}, 012330 (2007)}.

\bibitem{SerafiniPRL2010}  A. Serafini and S. Mancini, Determination of Maximal Gaussian Entanglement Achievable by Feedback-Controlled Dynamics, \href{https://doi.org/10.1103/PhysRevLett.104.220501}{Phys. Rev. Lett. \textbf{104}, 220501 (2010)}.

\bibitem{DohertyPRA1999} A. C. Doherty and K. Jacobs, Feedback control of quantum systems using continuous state estimation, \href{https://doi.org/10.1103/PhysRevA.60.2700}{Phys. Rev. A \textbf{60}, 2700 (1999)}.

\bibitem{Conangla2019}   G. P. Conangla, F. Ricci, M. T. Cuairan, A. W. Schell, N. Meyer, and R. Quidant, Optimal Feedback Cooling of a Charged Levitated Nanoparticle with Adaptive Control, \href{https://doi.org/10.1103/PhysRevLett.122.223602}{Phys. Rev. Lett. \textbf{122}, 223602 (2019)}.

\bibitem{MagriniNature2021} F. Tebbenjohanns, M. L. Mattana, M. Rossi, M. Frimmer, and L. Novotny, Quantum control of a nanoparticle optically levitated in cryogenic free space, \href{https://doi.org/10.1038/s41586-021-03617-w}{Nature (London) \textbf{595}, 378 (2021)}.

\bibitem{RosenzweigNature2021} L. Magrini, P. Rosenzweig, C. Bach, A. Deutschmann-Olek, S. G. Hofer, S. Hong, N. Kiesel, A. Kugi, and M. Aspelmeyer, Real-time optimal quantum control of mechanical motion at room temperature, \href{https://doi.org/10.1038/s41586-021-03602-3}{Nature (London) \textbf{595}, 373 (2021)}.

\bibitem{Lukin2004} J. K. Stockton, R. van Handel, and H. Mabuchi, Deterministic Dicke-state preparation with continuous measurement and control, \href{https://doi.org/10.1103/PhysRevA.70.022106}{Phys. Rev. A
\textbf{70}, 022106 (2004)}.


\bibitem{Dong2020} D. Dong, and I. R. Petersen, \emph{Learning and Robust Control in Quantum Technology}, (Springer Nature, Switzerland, 2023).

 \bibitem{Rabitz2000sc}  H. Rabitz, R. de Vivie-Riedle, M. Motzkus, and K. Kompa, Whither
the future of controlling quantum phenomena?, \href{https://www.science.org/doi/abs/10.1126/science.288.5467.824}{Science \textbf{288}, 824 (2000)}.

\bibitem{JudsonPRL1992} R. S. Judson and H. Rabitz, Teaching Lasers to Control Molecules, \href{https://doi.org/10.1103/PhysRevLett.68.1500}{Phys. Rev. Lett. \textbf{68}, 1500 (1992)}.

\bibitem{TwamleyPRA2009} J.-H. Sch\"{o}nfeldt, J. Twamley, and S. Rebi\'{c}, Optimized control of Stark-shift-chirped rapid adiabatic passage in a $\lambda$-type three-level system, \href{https://doi.org/10.1103/PhysRevA.80.043401}{Phys. Rev. A \textbf{80}, 043401 (2009)}.

\bibitem{Zahedinejad2015} E. Zahedinejad, J. Ghosh, and B. C. Sanders, High-Fidelity Single-Shot Toffoli Gate via Quantum Control, \href{https://doi.org/10.1103/PhysRevLett.114.200502}{Phys. Rev. Lett. \textbf{114}, 200502 (2015)}.

\bibitem{WuIE2017} D. Dong, X. Xing, H. Ma, C. Chen, Z. Liu, and H. Rabitz, Learning-Based Quantum Robust Control: Algorithm, Applications, and Experiments, \href{https://ieeexplore.ieee.org/abstract/document/8759071}{IEEE Trans. Cybernet. \textbf{50}, 3581 (2020)}.

\bibitem{NiuNPJ2019} M. Y. Niu, S. Boixo, V. N. Smelyanskiy, and H. Neven, Universal quantum control through deep reinforcement learning, \href{https://doi.org/10.1038/s41534-019-0141-3}{npj Quantum Inf. \textbf{5}, 33 (2019)}.

\bibitem{Sekatski2018} F. Fr\"{o}wis, P. Sekatski, W. D\"{u}r, N. Gisin, and N. Sangouard,
Macroscopic quantum states: Measures, fragility, and implementations, \href{https://doi.org/10.1103/RevModPhys.90.025004}{Rev. Mod. Phys. \textbf{90}, 025004 (2018)}.

\bibitem{LiaoPRL2016} J.-Q. Liao and L. Tian, Macroscopic Quantum Superposition in Cavity Optomechanics, \href{https://doi.org/10.1103/PhysRevLett.116.163602}{Phys. Rev. Lett. \textbf{116}, 163602 (2016)}.

\bibitem{CavesRMP1980}  C. M. Caves, K. S. Thorne, R. W. P. Drever, V. D. Sandberg, and M. Zimmermann, On the measurement of a weak classical force coupled to a quantum-mechanical oscillator. I. Issues of principle, \href{https://doi.org/10.1103/RevModPhys.52.341}{Rev. Mod. Phys. \textbf{52}, 341 (1980)}.

\bibitem{Giovannetti2004} V. Giovannetti, S. Lloyd, and L. Maccone, Quantum-Enhanced Measurements: Beating the Standard Quantum Limit, \href{https://www.science.org/doi/abs/10.1126/science.1104149}{Science \textbf{306}, 1330 (2004)}.

\bibitem{MetcalfeAPR2014} M. Metcalfe, Applications of cavity optomechanics, \href{https://doi.org/10.1063/1.4896029}{Appl. Phys. Rev. \textbf{1}, 031105 (2014)}.

\bibitem{GiganN2006} S. Gigan, H. R. Boehm, M. Paternostro, F. Blaser, G. Langer, J. B. Hertzberg, K. C. Schwab, D. Baeuerle, M. Aspelmeyer, and A. Zeilinger, Self-cooling of a micromirror by radiation pressure, \href{https://doi.org/10.1038/nature05273}{Nature (London) \textbf{444}, 67 (2006)}.

\bibitem{ArcizetN2006}  O. Arcizet, P.-F. Cohadon, T. Briant, M. Pinard, and A. Heidmann, Radiation-pressure cooling and optomechanical instability of a micromirror, \href{https://doi.org/10.1038/nature05244}{Nature (London) \textbf{444}, 71 (2006)}.

\bibitem{KlecknerN2006} D. Kleckner and D. Bouwmeester, Sub-kelvin optical cooling of a micromechanical resonator, \href{https://doi.org/10.1038/nature05231}{Nature (London) \textbf{444}, 75 (2006)}.

\bibitem{Wilson-Rae2007PRL} I. Wilson-Rae, N. Nooshi, W. Zwerger, and T. J.
Kippenberg, Theory of Ground State Cooling of a Mechanical Oscillator Using
Dynamical Backaction, \href{https://doi.org/10.1103/PhysRevLett.99.093901}{Phys. Rev. Lett. \textbf{99}, 093901 (2007)}.

\bibitem{Marquardt2007PRL} F. Marquardt, J. P. Chen, A. A. Clerk, and S. M.
Girvin, Quantum Theory of Cavity-Assisted Sideband Cooling of Mechanical
Motion, \href{https://doi.org/10.1103/PhysRevLett.99.093902}{Phys. Rev. Lett. \textbf{99}, 093902 (2007)}.

\bibitem{Dobrindt2008PRL} J. M. Dobrindt, I. Wilson-Rae, and T. J.
Kippenberg, Parametric Normal-Mode Splitting in Cavity Optomechanics, \href{https://doi.org/10.1103/PhysRevLett.101.263602}{Phys.
Rev. Lett. \textbf{101}, 263602 (2008)}.

\bibitem{Genes200877PRA}  C. Genes, D. Vitali, P. Tombesi, S. Gigan, and M. Aspelmeyer,
Ground-state cooling of a micromechanical oscillator: Comparing cold damping and cavity-assisted cooling schemes, \href{https://doi.org/10.1103/PhysRevA.77.033804}{Phys.
Rev. A \textbf{77}, 033804 (2008)}.

\bibitem{Chan2011Nature} J. Chan, T. P. M. Alegre, A. H. Safavi-Naeini, J.
T. Hill, A. Krause, S. Gr\"{o}blacher, M. Aspelmeyer, and O. Painter, Laser
cooling of a nanomechanical oscillator into its quantum ground state, \href{https://doi.org/10.1038/nature10461}{Nature
(London) \textbf{478}, 89 (2011)}.

\bibitem{Teufel2011Nature} J. D. Teufel, T. Donner, D. Li, J. W. Harlow, M.
S. Allman, K. Cicak, A. J. Sirois, J. D. Whittaker, K. W. Lehnert, and R. W.
Simmonds, Sideband cooling of micromechanical motion to the quantum ground
state, \href{https://doi.org/10.1038/nature10261}{Nature (London) \textbf{475}, 359 (2011)}.

\bibitem{Rossi2018Nature} M. Rossi, D. Mason, J. Chen, Y. Tsaturyan, and A.
Schliesser, Measurement-based quantum control of mechanical motion, \href{https://doi.org/10.1038/s41586-018-0643-8}{Nature
(London) \textbf{563}, 53 (2018)}.

\bibitem{Sommer2019PRL} C. Sommer and C. Genes, Partial Optomechanical
Refrigeration via Multimode Cold-Damping Feedback, \href{https://doi.org/10.1103/PhysRevLett.123.203605}{Phys. Rev. Lett. \textbf{%
123}, 203605 (2019)}.

\bibitem{LaiPRA2020} D.-G. Lai, J.-F. Huang, X.-L. Yin, B.-P. Hou, W. Li, D.
Vitali, F. Nori, and J.-Q. Liao, Nonreciprocal ground-state
cooling of multiple mechanical resonators, \href{https://doi.org/10.1103/PhysRevA.102.011502}{Phys. Rev. A \textbf{102}, 011502(R) (2020)}.

\bibitem{LiuPRA2022} Y.-H. Liu, X.-L. Yin, J.-F. Huang, and J.-Q. Liao,
Accelerated ground-state cooling of an optomechanical resonator via
shortcuts to adiabaticity, \href{https://doi.org/10.1103/PhysRevA.105.023504}{Phys. Rev. A \textbf{105}, 023504 (2022)}.

\bibitem{Vitali2007PRL} D. Vitali, S. Gigan, A. Ferreira, H. R. B\"{o}hm, P.
Tombesi, A. Guerreiro, V. Vedral, A. Zeilinger, and M. Aspelmeyer,
Optomechanical Entanglement between a Movable Mirror and a Cavity Field,
\href{https://doi.org/10.1103/PhysRevLett.98.030405}{Phys. Rev. Lett. \textbf{98}, 030405 (2007)}.

\bibitem{Genes2008PRA} C. Genes, A. Mari, P. Tombesi, and D. Vitali, Robust entanglement of a micromechanical resonator with output optical fields, \href{https://doi.org/10.1103/PhysRevA.78.032316}{Phys. Rev. A \textbf{78}, 032316 (2008)}.

\bibitem{Palomaki2013Science} T. A. Palomaki, J. D. Teufel, R. W. Simmonds,
and K. W. Lehnert, Entangling Mechanical Motion with Microwave Fields,
\href{https://www.science.org/doi/abs/10.1126/science.1244563}{Science \textbf{342}, 710 (2013)}.

\bibitem{Riedinger2016Nature} R. Riedinger, S. Hong, R. A. Norte, J. A.
Slater, J. Shang, A. G. Krause, V. Anant, M. Aspelmeyer, and S. Gr\"{o}blacher,
Non-classical correlations between single photons and phonons from a
mechanical oscillator, \href{https://doi.org/10.1038/nature16536}{Nature (London) \textbf{530}, 313 (2016)}.

\bibitem{Hong2017Sci} S. Hong, R. Riedinger, I. Marinkovi\'{c}, A. Wallucks, S. G. Hofer, R. A. Norte, M. Aspelmeyer, and S. Gr\"{o}blacher, Hanbury Brown and Twiss interferometry of single phonons from an optomechanical resonator, \href{https://www.science.org/doi/10.1126/science.aan7939}{Science \textbf{358}, 203 (2017)}.

\bibitem{Ho2018PRL} M. Ho, E. Oudot, J.-D. Bancal, and N. Sangouard,
Witnessing Optomechanical Entanglement with Photon Counting, \href{https://doi.org/10.1103/PhysRevLett.121.023602}{Phys. Rev.
Lett. \textbf{121}, 023602 (2018)}.

\bibitem{RiedingerNA2018} R. Riedinger, A. Wallucks, I. Marinkovi\'{c}, C.L\"{o}schnauer, M. Aspelmeyer, S. Hong, and S. Gr\"{o}blacher, Remote quantum entanglement between two micromechanical oscillators, \href{https://doi.org/10.1038/s41586-018-0036-z}{Nature (London) \textbf{556}, 473 (2018)}.

\bibitem{KorppiN2018} C. F. Ockeloen-Korppi, E. Damsk\"{a}gg, J. M. Pirkkalainen, M. Asjad, A. A. Clerk, F. Massel, M. J. Woolley, and M. A. Sillanp\"{a}\"{a}, Stabilized entanglement of massive mechanical oscillators, \href{https://doi.org/10.1038/s41586-018-0038-x}{Nature (London) \textbf{556}, 478 (2018)}.

\bibitem{Yu2020Nature} H. Yu, L. McCuller, M. Tse, N. Kijbunchoo, L.
Barsotti, N. Mavalvala, and L. S. Collaboration, Quantum correlations
between light and the kilogram-mass mirrors of LIGO, \href{https://doi.org/10.1038/s41586-020-2420-8}{Nature (London) \textbf{583}, 43 (2020)}.

\bibitem{Jiao2020PRL} Y.-F. Jiao, S.-D. Zhang, Y.-L. Zhang, A. Miranowicz,
L.-M. Kuang, and H. Jing, Nonreciprocal Optomechanical Entanglement against
Backscattering Losses, \href{https://doi.org/10.1103/PhysRevLett.125.143605}{Phys. Rev. Lett. \textbf{125}, 143605 (2020)}.

\bibitem{Lai2022PRL} D.-G. Lai, J.-Q. Liao, A. Miranowicz, and F. Nori, Noise Tolerant Optomechanical Entanglement via Synthetic Magnetism, \href{https://doi.org/10.1103/PhysRevLett.129.063602}{Phys. Rev. Lett. \textbf{129}, 063602 (2022)}.

\bibitem{Greiner1996sp} W. Greiner and J. Reinhardt, \emph{Field Quantization%
} (Springer, Berlin, 1996).

\bibitem{Henderson1981} H. V. Henderson and S. R. Searle, The
vec-permutation matrix, the vec operator and Kronecker products: a review,
\href{https://doi.org/10.1080/03081088108817379}{Lin. Mult. Alg. \textbf{9},
271 (1981)}.

\bibitem{Scully1997} M. O. Scully and M. S. Zubairy, \emph{Quantum Optics}
(Cambridge University Press, Cambridge, England, 1997).

\bibitem{Agarwal2013} G. S. Agarwal, \emph{Quantum Optics} (Cambridge
University Press, Cambridge, England, 2013).

\bibitem{MariPRL2009} A. Mari and J. Eisert, Gently Modulating Optomechanical Systems, \href{https://doi.org/10.1103/PhysRevLett.103.213603}{Phys. Rev. Lett. \textbf{103}, 213603 (2009)}.

\bibitem{VannerPNAS2011} M. R. Vanner, I. Pikovski, G. D. Cole, M. S. Kim, C. Brukner, K. Hammerer, G. J. Milburn, and M. Aspelmeyer, Pulsed quantum optomechanics, \href{https://doi.org/10.1073/pnas.1105098108}{Proc. Natl. Acad. Sci. \textbf{108}, 16182 (2011)}.

\bibitem{HoferPRA2011} S. G. Hofer, W. Wieczorek, M. Aspelmeyer, and K. Hammerer, Quantum entanglement and teleportation in pulsed cavity optomechanics, \href{https://doi.org/10.1103/PhysRevA.84.052327}{Phys. Rev. A \textbf{84}, 052327 (2011)}.

 \bibitem{LiaoPRA201184}  J.-Q. Liao and C. K. Law, Cooling of a mirror in cavity optomechanics with a chirped pulse, \href{https://doi.org/10.1103/PhysRevA.84.053838}{Phys. Rev. A \textbf{84}, 053838 (2011)}.

\bibitem{MachnesPRL2012}  S. Machnes, J. Cerrillo, M. Aspelmeyer, W. Wieczorek, M. B. Plenio, and A. Retzker, Pulsed Laser Cooling for Cavity Optomechanical Resonators, \href{https://doi.org/10.1103/PhysRevLett.108.153601}{Phys. Rev. Lett. \textbf{108}, 153601 (2012)}.

 \bibitem{VannerNC2013} M. R. Vanner, J. Hofer, G. D. Cole, and M. Aspelmeyer, Cooling-by-measurement and mechanical state tomography via pulsed optomechanics, \href{https://doi.org/10.1038/ncomms3295}{Nat. Commun. \textbf{4}, 2295 (2013)}.

\bibitem{Palomaki2013} T. A. Palomaki, J. W. Harlow, J. D. Teufel, R. W. Simmonds, and K. W. Lehnert, Coherent state transfer between itinerant microwave fields and a mechanical oscillator, \href{https://doi.org/10.1038/nature11915}{Nature (London) \textbf{495}, 210 (2013)}.

 \bibitem{MeenehanPRX2015}  S. M. Meenehan, J. D. Cohen, G. S. MacCabe, F. Marsili, M. D. Shaw, and O. Painter, Pulsed Excitation Dynamics of an Optomechanical Crystal Resonator near Its Quantum Ground State of Motion, \href{https://doi.org/10.1103/PhysRevX.5.041002}{Phys. Rev. X \textbf{5}, 041002 (2015)}.

\bibitem{MacCabe2020Sc} G. S. MacCabe, H. J. Ren, J. Luo, J. D. Cohen, H. Y. Zhou, A. Sipahigil, M. Mirhosseini, and O. Painter, Nano acoustic resonator with ultralong phonon life time, \href{https://www.science.org/doi/10.1126/science.abc7312}{Science \textbf{370}, 840 (2020)}.

  \bibitem{FedoseevPRL2021}  V. Fedoseev, F. Luna, I. Hedgepeth, W. L\"{o}ffler, and D. Bouwmeester, Stimulated Raman Adiabatic Passage in Optomechanics, \href{https://doi.org/10.1103/PhysRevLett.126.113601}{Phys. Rev. Lett. \textbf{126}, 113601 (2021)}.

\bibitem{He2017PRL} B. He, L. Yang, Q. Lin, and M. Xiao, Radiation Pressure Cooling as a Quantum Dynamical Process, \href{https://doi.org/10.1103/PhysRevLett.118.233604}{Phys. Rev. Lett. \textbf{118}, 233604 (2017)}.

\bibitem{TrianaPRL2016} J. F. Triana, A. F. Estrada, and L. A. Pach\'{o}n, Ultrafast Optimal Sideband Cooling under Non-Markovian Evolution, \href{https://doi.org/10.1103/PhysRevLett.116.183602}{Phys. Rev. Lett. 116, 183602 (2016)}.

\bibitem{AdessoPRA2004} G. Adesso, A. Serafini, and F. Illuminati, Extremal entanglement and mixedness in continuous variable systems, \href{https://doi.org/10.1103/PhysRevA.70.022318}{Phys. Rev. A \textbf{70}, 022318 (2004)}.

\bibitem{PlenioPRL205} M. B. Plenio, Logarithmic Negativity: A Full Entanglement
Monotone That Is Not Convex, \href{https://doi.org/10.1103/PhysRevLett.95.090503}{Phys. Rev. Lett. \textbf{95}, 090503 (2005)}.


\bibitem{LiuNPJquan} Y. Liu, Q. Liu, H. Sun, M. Chen, S. Wang, and T. Li, Coherent memory for microwave photons based on long-lived mechanical excitations, \href{https://doi.org/10.1038/s41534-023-00749-x}{npj Quantum Inf. \textbf{9}, 80 (2023)}.

\bibitem{LHNp2019} L. Hu, Y. Ma, W. Cai, X. Mu, Y. Xu, W. Wang, Y. Wu, H. Wang, Y. P. Song, C. L. Zou, S. M. Girvin, L-M. Duan, and L. Sun, Quantum error correction and universal gate set operation on a binomial bosonic logical qubit, \href{https://doi.org/10.1038/s41567-018-0414-3}{Nat. Phys. \textbf{15}, 503(2019)}.

\bibitem{Meyer2019} N. Meyer, A. R. Sommer, P. Mestres, J. Gieseler, V. Jain, L. Novotny, and R. Quidant, Resolved-Sideband Cooling of a Levitated Nanoparticle in the Presence of Laser Phase Noise, \href{https://doi.org/10.1103/PhysRevLett.123.153601}{Phys. Rev. Lett. \textbf{123}, 153601 (2019)}.

\bibitem{ShomroniPRL} L. Qiu, I. Shomroni, P. Seidler, and T. J. Kippenberg, Laser Cooling of a Nanomechanical Oscillator to Its Zero-Point Energy,  \href{https://doi.org/10.1103/PhysRevLett.124.173601}{Phys. Rev. Lett. \textbf{124}, 173601 (2020).}

\bibitem{Zeng2022} Y. X. Zeng, B. Xiong, and C. Li, Suppressing laser phase noise in an optomechanical system, \href{https://doi.org/10.1007/s11467-021-1097-2}{Front. Phys. \textbf{17}, 12503 (2022)}.

\bibitem{LPRA2008} L. Di\'{o}si, Laser linewidth hazard in optomechanical cooling, \href{https://doi.org/10.1103/PhysRevA.78.021801}{Phys. Rev. A \textbf{78}, 021801(R) (2008)}.

\bibitem{Eichenfield2009} M. Eichenfield, R. Camacho, J. Chan, K. J. Vahala, and O. Painter, A picogram- and nanometre-scale photonic-crystal optomechanical cavity, \href{https://doi.org/10.1038/nature08061}{Nature (London) \textbf{459}, 550 (2009)}.

\bibitem{ChanNT2009}  M. Eichenfield, J. Chan, R. M. Camacho, K. J. Vahala, and O. Painter, Optomechanical crystals, \href{https://doi.org/10.1038/nature08524}{Nature (London) \textbf{462}, 78 (2009)}.

 \bibitem{HertzbergNP2010}  J. B. Hertzberg, T. Rocheleau, T. Ndukum, M. Savva, A. A. Clerk, and K. C. Schwab, Back-action-evading measurements of nanomechanical motion, \href{https://doi.org/10.1038/nphys1479}{Nat. Phys. \textbf{6}, 213 (2010)}.

 \bibitem{MasselN2011} F. Massel, T. T. Heikkil\"{a}, J.-M. Pirkkalainen, S. U. Cho, H. Saloniemi, P. J. Hakonen, and M. A. Sillanp\"{a}\"{a}, Microwave amplification with nanomechanical resonators, \href{https://doi.org/10.1038/nature10628}{Nature (London) \textbf{480}, 351 (2011)}.

\bibitem{MasselNC2012} F. Massel, S. U. Cho, J.-M. Pirkkalainen, P. J. Hakonen, T. T. Heikkil\"{a}, and M. A. Sillanp\"{a}\"{a}, Multimode circuit optomechanics near the quantum limit, \href{https://doi.org/10.1038/ncomms1993}{Nat. Commun. \textbf{3}, 987 (2012)}.

\bibitem{ParkNP2009} Y. S. Park and H. Wang, Resolved-sideband and cryogenic cooling of an optomechanical resonator, \href{https://doi.org/10.1038/nphys1303}{Nat. Phys. \textbf{5}, 489 (2009)}.

\bibitem{SchliesserNP2009} A. Schliesser, O. Arcizet, R. Rivi\`{e}re, G. Anetsberger, and T. J. Kippenberg, Resolved-sideband cooling and position measurement of a micromechanical oscillator close to the Heisenberg uncertainty limit, \href{https://doi.org/10.1038/nphys1304}{Nat. Phys. \textbf{5}, 509 (2009)}.

\bibitem{VerhagenNature2012} E. Verhagen, S. Del\'{e}glise, S. Weis, A. Schliesser, and T. J. Kippenberg, Quantum-coherent coupling of a mechanical oscillator to an optical cavity mode, \href{https://doi.org/10.1038/nature10787}{Nature (London) \textbf{482}, 63 (2012)}.

\bibitem{HammererNature2009} S. Gr\"{o}blacher, K. Hammerer, M. R. Vanner, and M.
Aspelmeyer, Observation of strong coupling between a micromechanical resonator and an optical cavity field, \href{https://doi.org/10.1038/nature08171}{Nature (London) \textbf{460}, 724 (2009)}.


\end{thebibliography}
\end{document}